\documentclass[sigconf, nonacm]{acmart}

\AtBeginDocument{%
  \providecommand\BibTeX{{%
    Bib\TeX}}}

\usepackage{hyperref}
\usepackage{xcolor}
\usepackage{algorithm}
\usepackage{xspace}
\usepackage{bbding}
\usepackage{pifont}
\usepackage{multirow}

\usepackage{threeparttable}
\usepackage{algorithmicx}
\usepackage{algpseudocode}

\usepackage{fancyhdr}

\newcommand{\bheading}[1]{{\vspace{4pt}\noindent{\textbf{#1}}}}

\newcommand{\kl}[1]{\textcolor{pink}{KL: #1}}

\newcounter{note}[section]
\renewcommand{\thenote}{\thesection.\arabic{note}}
\newcommand{\zhe}[1]{\refstepcounter{note}{\bf\textcolor{purple}{$\ll$Zhe~\thenote: #1$\gg$}}}

\newcommand{\sys}{\mbox{\textsc{SyzParam}}\xspace}

\newenvironment{packeditemize}{
    \begin{list}{$\scriptstyle\bullet$}{
        \setlength{\labelwidth}{8pt}
        \setlength{\itemsep}{0pt}
        \setlength{\leftmargin}{\labelwidth}
        \addtolength{\leftmargin}{\labelsep}
        \setlength{\parindent}{0pt}
        \setlength{\listparindent}{\parindent}
        \setlength{\parsep}{2pt}
        \setlength{\topsep}{3pt}}}{
    \end{list}
}

\def\Snospace~{\S{}}

\newcommand{\patrick}[1]{{\bf\textcolor{red}{$\lceil$R\thenote@Patrick: {\sf #1}$\rfloor$}}}

\begin{document}

\title{\sys: Incorporating Runtime Parameters into Kernel Driver Fuzzing}

\author{Yue Sun}
\affiliation{%
  \institution{SKLP, Institute of Computing Technology, CAS \& University of Chinese Academy of Sciences}
  \city{Beijing}
  \country{China}}
\email{sunyue163@mails.ucas.ac.cn}
\author{Yan Kang}
\affiliation{%
  \institution{SKLP, Institute of Computing Technology, CAS \& University of Chinese Academy of Sciences}
  \city{Beijing}
  \country{China}}
\email{kangyan@ict.ac.cn}
\author{Chenggang Wu}
\authornote{Corresponding author.}
\affiliation{%
  \institution{SKLP, Institute of Computing Technology, CAS \& University of Chinese Academy of Sciences \& Zhongguancun Laboratory}
  \city{Beijing}
  \country{China}}
\email{wucg@ict.ac.cn}
\author{Kangjie Lu}
\affiliation{%
  \institution{University of Minnesota}
  \city{Minneapolis}
  \country{USA}}
\email{kjlu@umn.edu}
\author{Jiming Wang}
\affiliation{%
  \institution{SKLP, Institute of Computing Technology, CAS \& University of Chinese Academy of Sciences}
  \city{Beijing}
  \country{China}}
\email{wangjiming21@mails.ucas.ac.cn}
\author{Xingwei Li}
\affiliation{%
  \institution{Information Engineering University}
  \city{Beijing}
  \country{China}}
\email{xrivendell7@outlook.com}
\author{Yuhao Hu}
\affiliation{%
  \institution{SKLP, Institute of Computing Technology, CAS \& University of Chinese Academy of Sciences}
  \city{Beijing}
  \country{China}}
\email{huyuhao19f@ict.ac.cn}
\author{Jikai Ren}
\affiliation{%
  \institution{SKLP, Institute of Computing Technology, CAS \& University of Chinese Academy of Sciences}
  \city{Beijing}
  \country{China}}
\email{renjikai22@mails.ucas.ac.cn}
\author{Yuanming Lai}
\affiliation{%
  \institution{SKLP, Institute of Computing Technology, CAS}
  \city{Beijing}
  \country{China}}
\email{laiyuanming19b@ict.ac.cn}
\author{Mengyao Xie}
\affiliation{%
  \institution{SKLP, Institute of Computing Technology, CAS}
  \city{Beijing}
  \country{China}}
\email{xiemengyao@ict.ac.cn}
\author{Zhe Wang}
\affiliation{%
  \institution{SKLP, Institute of Computing Technology, CAS \& Zhongguancun Laboratory}
  \city{Beijing}
  \country{China}}
\email{wangzhe12@ict.ac.cn}

\renewcommand{\shortauthors}{Sun et al.}

\settopmatter{printfolios=true,printccs=false,printacmref=false}

\begin{abstract}
Under the monolithic architecture of the Linux kernel, all its components operate within the same address space. Notably, device drivers constitute over half of the kernel codebase yet are particularly prone to bugs, incurring significant risks. Therefore, exploring vulnerabilities in drivers is critical for ensuring kernel security. Extensive research has been done to fuzz kernel drivers through system calls and hardware interrupts to trigger execution errors. However, significant portions of the codebase remain uncovered despite rigorous testing efforts.

Through a comprehensive study of the Linux Kernel Device Model (LKDM), we identified that the execution of device drivers is influenced not only by the existing input spaces mentioned above but also by runtime parameters, including device attributes and kernel module parameters. Our analysis reveals that large portions of the uncovered code are masked by these parameters, which are exposed to the userspace through a specialized virtual file system known as \texttt{sysfs}. Furthermore, adjacent devices interconnected within the same device tree also impact drivers' behavior.

This paper introduces a novel fuzzing framework, \sys, which incorporates runtime parameters into the fuzzing process. Achieving this objective requires addressing several key challenges, including valid value extraction, inter-device relation construction, and fuzz engine integration. By inspecting the data structures and functions associated with the LKDM, our tool can extract runtime parameters across various drivers through static analysis. Additionally, \sys collects inter-device relations and identifies associations between runtime parameters and drivers. Furthermore, \sys proposes a novel mutation strategy, which leverages these relations and prioritizes parameter modification during related driver execution. Our evaluation demonstrates that \sys outperforms existing fuzzing works in driver code coverage and bug-detection capabilities. To date, we have identified 30 unique bugs in the latest kernel upstreams, with 20 confirmed and 14 patched into the mainline kernel, including 9 CVEs.

\end{abstract}



\received{20 February 2007}
\received[revised]{12 March 2009}
\received[accepted]{5 June 2009}

\maketitle
\thispagestyle{fancy}
\cfoot{\thepage}
\pagestyle{fancy}
\cfoot{\thepage}

\section{Introduction}\label{Introduction}
As one of the most popular operating systems, Linux runs on servers and mobile devices worldwide. Due to its monolithic architecture, all drivers run in the same address space, which makes kernel security very vulnerable---a flaw in the driver could be exploited to compromise the whole kernel. According to statistics, over half of the kernel code comes from device drivers~\cite{kadav2012understanding}, and 27\%-54\% of Linux kernel CVE reports in recent 5 years are related to drivers according to previous research~\cite{wu2023devfuzz}. Therefore, discovering vulnerabilities in drivers is crucial to improving kernel security.

Fuzzing is one of the most popular methods to find vulnerabilities in software because of its effectiveness and scalability. In 2017, Google announced the release of Syzkaller~\cite{Syzkaller}, which has been used to fuzz kernels with different configurations. By the time of writing, Syzkaller has uncovered more than 7,500 vulnerabilities across different kernel versions. Recently, researchers have made numerous attempts to explore more vulnerabilities in Linux drivers, which can be categorized as hardware emulation and syscall enhancement. For hardware emulation, some literature has simulated absent hardware devices to bind and test drivers~\cite{zhao2022semantic,wu2023devfuzz}, while other studies have expanded the scope by introducing hardware interrupts as additional input space~\cite{hetzelt2021via,song2019periscope,ma2022printfuzz}.
For syscall enhancement, some researchers aim to automatically generate syscall descriptions for drivers utilizing kernel source code~\cite{corina2017difuze,sun2022ksg,hao2023syzdescribe}, and others focus on relations between syscalls~\cite{bulekov2023no,wang2021syzvegas,yang2023kernelgpt}. Additionally, some research integrates domain-specific knowledge into fuzzing~\cite{kim2022fuzzusb,peng2020usbfuzz,zhao2022statefuzz}. The initiatives above have all produced noteworthy outcomes, revealing kernel driver vulnerabilities from various angles.

Except for the existing input spaces discussed in previous works, we identify an additional factor that can significantly impact driver code execution. Under the current driver model, the Linux Kernel Device Model (LKDM), drivers expose runtime parameters to users via a specialized virtual file system (i.e., \texttt{sysfs}). These parameters can be broadly categorized into \emph{kernel module parameters} associated with drivers, and \emph{device attributes} linked to specific device instances. Through static analysis, we quantified the prevalence and influence of these parameters. Our findings reveal that the kernel includes over 2,000 types of runtime parameters, which directly impact more than 25,000 conditional statements (e.g., \texttt{if} and \texttt{switch} statements), dominating over 55,000 basic blocks. Compared to the 350,000 basic blocks that Syzkaller has covered through continuous fuzzing, these codes still take up 15.7\%. However, existing kernel driver fuzzing approaches neglect the impact of runtime parameters on driver code execution, thus failing to uncover these functionalities.

To gain a deeper understanding of runtime parameters, we carefully inspected historical bugs associated with runtime parameters. We observed that these parameters not only impact the execution of the driver to which they belong but can also impact other device drivers. By delving into the underlying data structures, we found that all devices are organized through the device tree, attaching to a parent device or bus during registration. Consequently, modification to one device will affect others through the device tree. Additionally, we identified that the timing of parameter modifications plays a critical role in triggering these bugs.

Given the findings above, incorporating runtime parameters in driver fuzzing is non-trivial. It faces three challenges: 
(1) \emph{How to pair runtime parameters and their corresponding values to pass input validation?} Although parameters can be modified by writing files under the \texttt{sysfs}, the kernel has strict logic to check and filter invalid inputs. Generating random values to write files will not change the value of the targeted parameters, thus failing to explore the new functionalities of drivers; 
(2) \emph{How to determine relations between runtime parameters and drivers?} Modifying the parameters alone cannot trigger new functionalities, but should be combined with the execution of the driver. However, there are over two thousand kinds of parameters. Randomly combining them with existing syscall descriptions will exponentially increase the search space, reducing the fuzzing efficiency significantly;
(3) \emph{How to integrate the runtime parameters into the fuzz engine and cooperate with other components in fuzzing?} Runtime parameter modification will have sustainable affection on the kernel driver execution, which will affect components in the fuzzing loop such as coverage calculation and seed minimization, leading to poor reproducibility.

To address the aforementioned challenges, we propose a two-stage driver fuzzing framework, \sys, which integrates runtime parameters into the driver fuzzing. Specifically, we design three targeted tasks to address these challenges systematically: \emph{First}, to enable the fuzzing engine to modify parameters dynamically, we implement a value collection task that identifies parameter filenames and determines values that can successfully pass input validation; \emph{Second}, to avoid the inefficiencies caused by random combinations of parameter modifications and driver executions, we perform a relation identification task. This task narrows the testing scope by focusing on inter-device connections and the relation between parameters and drivers; \emph{Third}, to ensure compatibility with functionalities in the fuzz engine, we propose new mutation strategies and leverage existing features to seamlessly integrate parameters into the fuzzing engine, maintaining orthogonality with other components in the fuzzing process.

We implemented \sys and evaluated it on the Linux v6.7. Overall, \sys can recover 1,243 types of device attributes, and 694 unique kernel module parameters. Compared with the fuzzing performance with Syzkaller, SyzDescribe, and Syzgen++ on 8 widely used device drivers on different busses, \sys achieves the highest edge coverage on 7 out of 8 drivers, with 32.57\% improvement in edge coverage compared to Syzkaller. Further study combines our work with existing work, and achieves the highest code coverage among all, proving that our coverage improvement is complementary to existing works. We also did an ablation study on vulnerability discovery capabilities and achieved a 34.5\% improvement in the number of unique crashes in a 120-hour trial in four rounds. \sys discovered 30 previously unknown bugs in total, among them 14 have been fixed into the upstream kernel, and 9 CVEs have been assigned so far.

In this paper, we make the following contributions:

\begin{packeditemize}
    \item \textbf{New findings on driver fuzzing.} We identify a long-neglected factor, runtime parameters, that can significantly impact driver code execution. Moreover, we conducted a thorough study of the LKDM, and pointed out that the inter-device relationship is the key factor to triggered the bug.
    \item \textbf{A novel driver fuzzing approach.} We propose a new driver fuzzing approach based on the findings above. We extend the ability of current kernel fuzzing tools by incorporating runtime parameters into driver fuzzing and utilizing the device relation tree to guide mutation strategies.
    \item \textbf{The implementation and evaluation of \sys.} We implement a fuzzing framework prototype of \sys and evaluate it comprehensively. Compared to related works, \sys could achieve higher code coverage among eight tested drivers. \sys has discovered 30 previously unknown vulnerabilities, of which 20 have been confirmed, 14 have been merged into upstream, and 9 CVEs have been assigned.
\end{packeditemize}

\section{Background}

In this section, we will first introduce the workflow of current kernel fuzzers and the recent progress in kernel driver fuzzing. Then, we will illustrate the current driver model in Linux, along with the runtime parameters it provides. An example will be given to demonstrate how runtime parameters influence the driver execution through a special virtual filesystem called \emph{sysfs}.

\begin{figure}[!ht]
\centering
\includegraphics[width=0.43\textwidth]{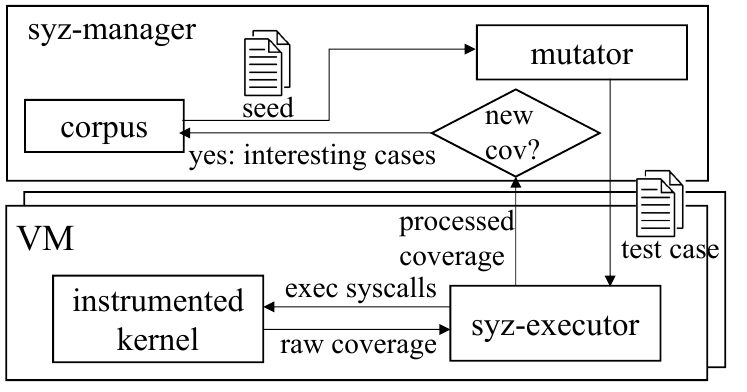}
\caption{The workflow of Syzkaller. The mutator in syz-manager is responsible for generating and mutating test cases, while syz-executor is responsible for executing syscalls and collecting coverage feedback.}
\label{figure1}
\Description{Workflow of Syzkaller, an open-sourced kernel fuzzer provided by Google.}
\end{figure}

\subsection{Linux Kernel Fuzzing}
Recently, many works have explored the Linux kernel vulnerabilities using the fuzzing technique~\cite{schumilo2017kafl, zhao2022statefuzz, jiang2022context, xumock}. Most were built on the popular kernel fuzzing framework Syzkaller~\cite{Syzkaller}. Syzkaller is a coverage-guided kernel fuzzer that uses system calls as inputs. The workflow of Syzkaller is illustrated in \autoref{figure1}. Syzkaller initializes multiple virtual machine (VM) instances, which are managed by a component called the \emph{syz-manager}, and executes a program named \emph{syz-executor} within each VM. The \emph{syz-manager} randomly selects a seed from the corpus, mutates it using predefined strategies, and generates test cases. The \emph{syz-executor} subsequently invokes the system calls specified in the test cases and records their code coverage after execution. When new coverage is detected, test cases that trigger the new coverage (referred to as ``interesting cases'') are minimized and incorporated into the corpus for future mutation.

\begin{figure}[!t]
\centering
\setlength{\abovecaptionskip}{0 cm}
\includegraphics[width=\columnwidth]{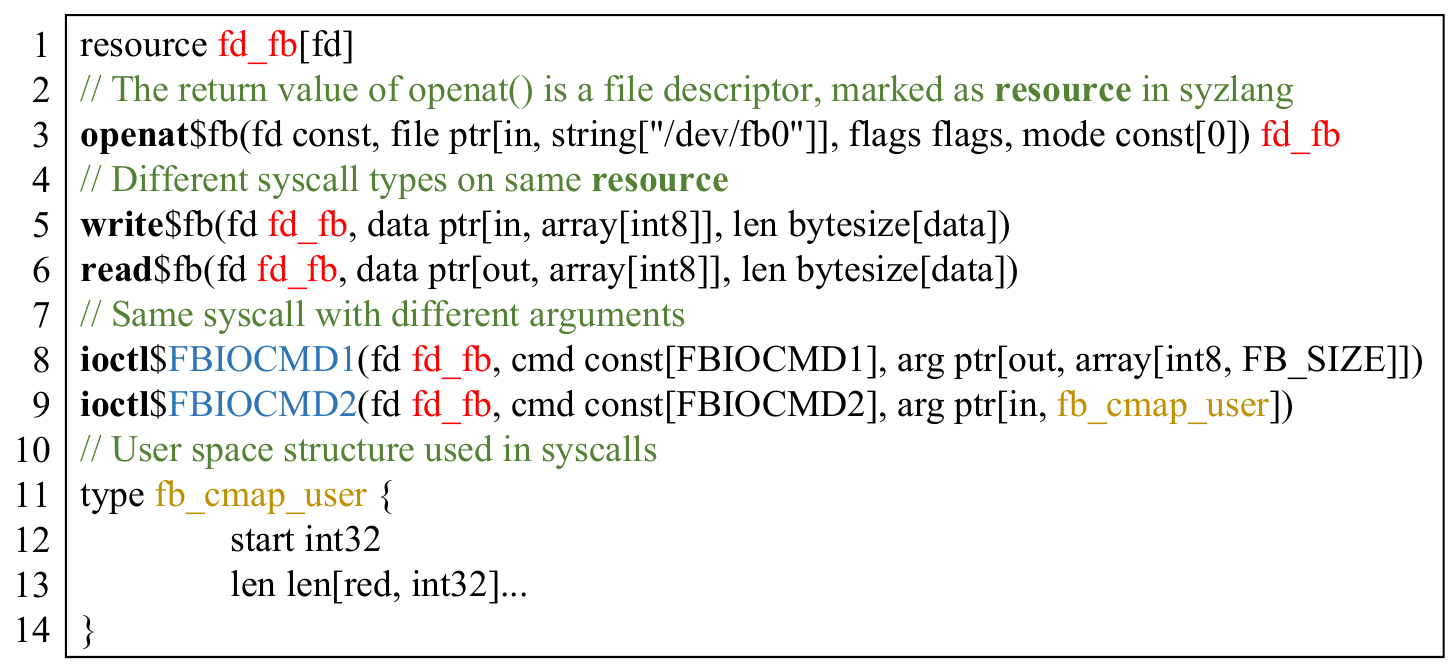}
\caption{An example of \emph{syzlang} provided by Syzkaller. \emph{Syzlangs} are declarative descriptions of syscall interfaces to manipulate programs, which will be used to generate, mutate, execute, minimize, serialize, and deserialize programs.}
\label{figure2}
\Description{Example of syzlangs, an intermediate representation used to describe system calls, provided by Syzkaller.}
\end{figure}

To generate complex system call sequences capable of triggering the execution of deep kernel code, Syzkaller introduced a syscall description language with type information, referred to as \emph{syzlang}~\cite{syzlang}. \emph{Syzlang} is a template that depicts the syscall types and their arguments. As shown in \autoref{figure2}, \emph{syzlang} supports various syscall types and argument types, including complex structure types (e.g., Lines 11-14). Given the wide variety of system calls and the fact that a single system call interface can accept different arguments and perform multiple functionalities, Syzkaller implements distinct \emph{syzlang} descriptions with different arguments (e.g., ioctl() in Lines 8-9). To establish data dependencies between \emph{syzlangs}, Syzkaller designates file descriptors as \texttt{resource} and establishes connections between \emph{syzlangs} through resource dependencies.

For complex logic that cannot be achieved by a single system call, Syzkaller provides an extension called \emph{pseudo-syscall}. These are C functions defined in the executor, which allows the executor to have code blocks to perform certain actions. They may also be used as more test-friendly wrappers for primitive syscalls. When Syzkaller generates a test case containing pseudo-syscalls, the executor invokes the predefined C functions.

\subsection{Existing Fuzzing Works on Linux Drivers}
Device drivers have long been and continue to be a significant source of defects and vulnerabilities in Linux kernels~\cite{shameli2021understanding}. Defects and vulnerabilities are an inherent part of the fast-growing and evolving driver codebase. Therefore, exploring device driver vulnerabilities is crucial to improving the security of the Linux kernel.


Existing research primarily focuses on effectively triggering the execution of drivers. Most approaches utilize the system call interface, as it provides the most accessible method for unprivileged users to initiate driver execution. Some studies aim to leverage static analysis and symbolic execution to extract driver-related system calls and their parameters from the kernel and automatically generate \emph{syzlangs}~\cite{chen2024syzgen++,hao2023syzdescribe,sun2022ksg,corina2017difuze}. Other works propose advanced mutation strategies to achieve higher coverage with existing \emph{syzlangs}~\cite{xumock,shen2022drifuzz,wang2021syzvegas}. Furthermore, some research highlights that interrupts can still serve as an attack surface~\cite{ma2022printfuzz}. Researchers try to simulate hardware and send arbitrary interrupts or return values to the kernel~\cite{wu2023devfuzz,ma2022printfuzz,hetzelt2021via,song2019periscope}, thus triggering driver execution.

\subsection{The Linux Kernel Device Model}

\begin{figure}[!t]
\centering
\includegraphics[width=\columnwidth]{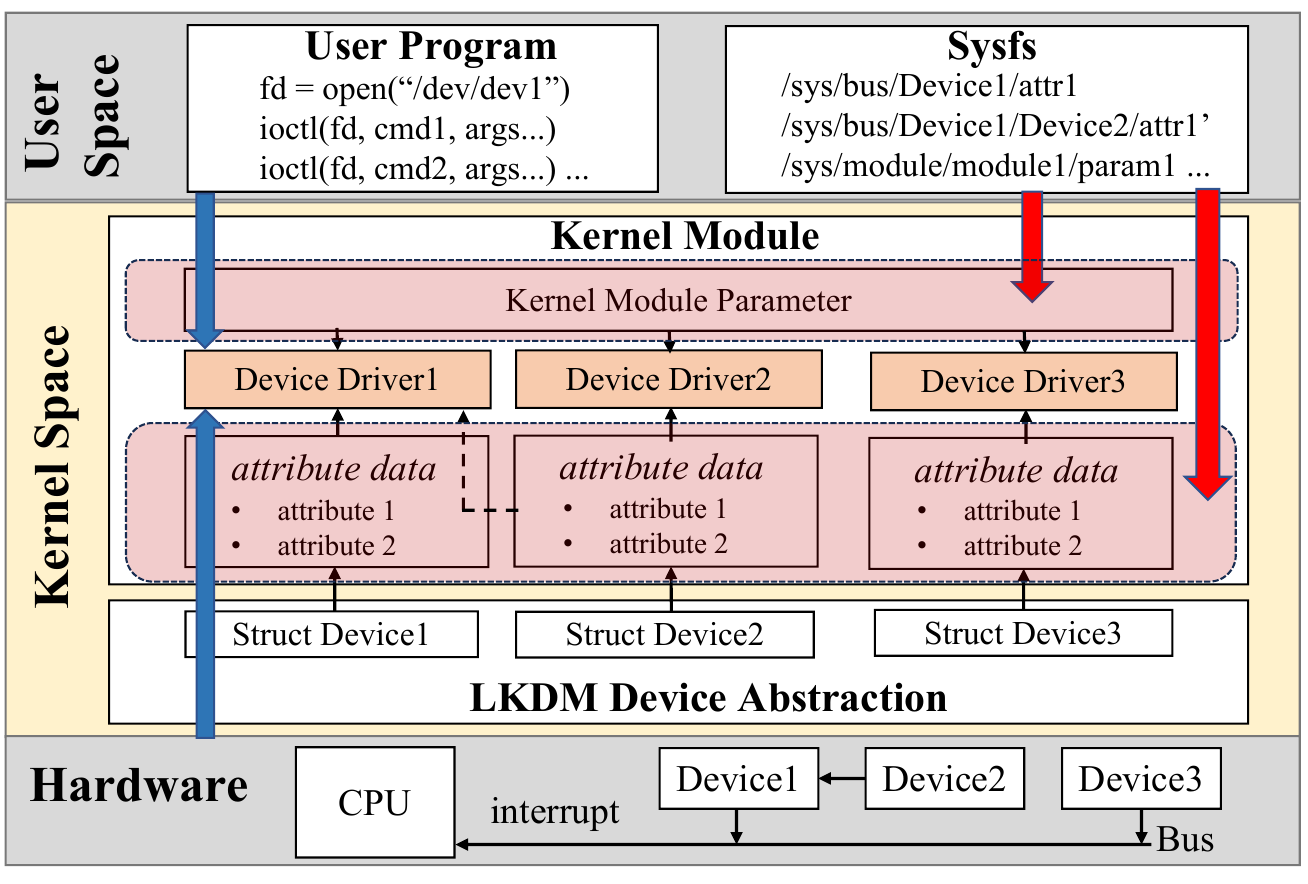}
\caption{An illustration of how runtime parameters affect kernel driver execution through sysfs. The red blocks represent the runtime parameters, and the orange blocks represent the device driver. Both user space and hardware can trigger code execution through blue arrows, and runtime parameters will influence the control flow through red arrows.}
\label{figure3}
\Description{Code that can be affected by sysfs, and the relationship between sysfs and other input spaces.}
\end{figure}

The Linux Kernel Device Model (LKDM) provides a common, uniform data model for describing buses and their corresponding devices~\cite{mochel2002linux}. Specifically, it implements device management from a bus-oriented perspective through a series of device-related data structures. Among these structures, we identify a series of variables that can influence the driver execution. We call them \emph{runtime parameters}, which can be broadly categorized into two types: \emph{kernel module parameters} and \emph{device attributes}.

\bheading{Kernel module parameters.}
Kernel module parameters are global variables defined within kernel modules, which can control driver behaviors. These variables are initialized when the kernel module is installed into the kernel. Since one or more device drivers may reside within the same kernel module, modifying the kernel module parameters might influence more than one driver's execution.

\bheading{Device attributes.}
Device attributes are driver- or bus-specific variables that record device status, and can also influence driver behaviors. It is worth pointing out that since devices are oriented in a tree-like structure physically, device attributes will not only affect the driver behavior of the current device, but also might influence the behavior of adjacent devices' drivers.

Linux employs a specialized virtual filesystem, known as \texttt{sysfs}, to expose kernel variables to user space. Due to the hierarchical nature of the bus-device structure, \texttt{sysfs} presents device structures and their relationships in the form of a file directory. 

We summarize how runtime parameters influence the driver execution through \texttt{sysfs}. As depicted in \autoref{figure3}, blue arrows indicate the inputs that trigger the code execution, including syscalls and interrupts, and red boxes represent the runtime parameters. Both kernel module parameters and device attributes are exposed to user space through \texttt{sysfs}, enabling users to configure the corresponding variables in the kernel to affect the driver's execution (through red arrows). \texttt{Sysfs} also reflects the underlying device connections. As shown in the bottom gray box of \autoref{figure3}, \texttt{Device 2} is a sub-device of \texttt{Device 1}, which is exported to user space by \texttt{sysfs} as hierarchical directories. Kernel module parameters reside in a separate folder. Modifying the \emph{device attributes} of \texttt{Device 2} may affect the driver execution of \texttt{Device 1} due to interconnected data structures.

\begin{figure}[!t]
\centering
\includegraphics[width=\columnwidth]{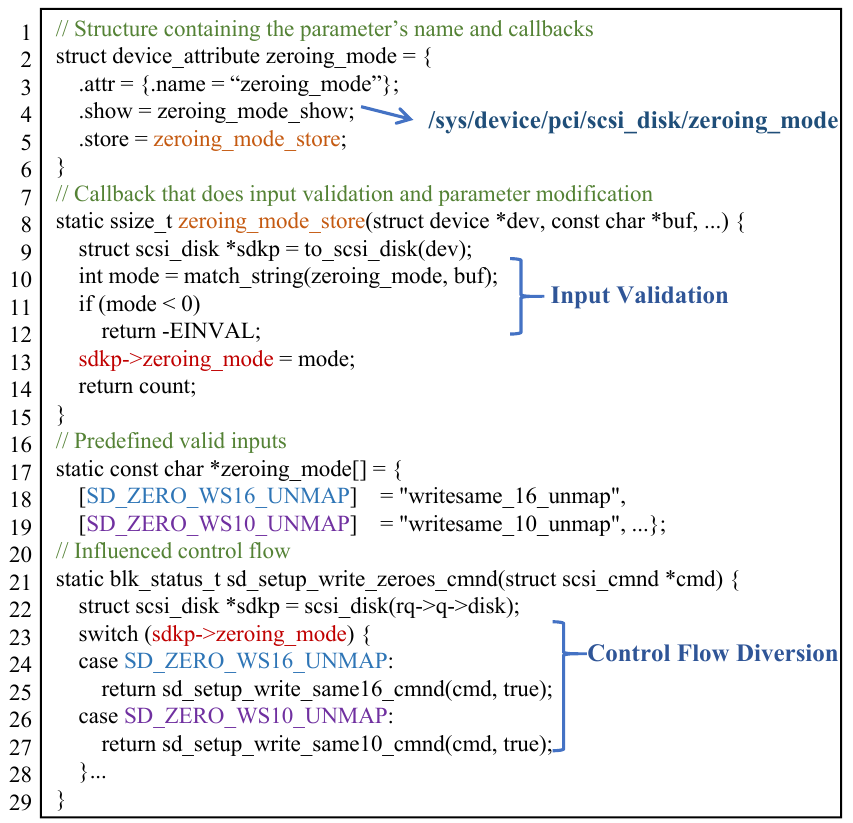}
\caption{An example of a runtime parameter and how it influences the driver execution.}
\label{figure4}
\Description{The parameter is exported to userspace by predefined code and callbacks.}
\end{figure}

We will give a brief example to illustrate how the runtime parameters get modified, and how they affect driver behaviors. In \autoref{figure4}, ``zeroing\_mode'' is one of the device attributes of scsi\_disk device. It is exported to user space through \texttt{sysfs} and uses a set of \emph{show/store} function callbacks. When the user modifies the file ``zeroing\_mode'', the \emph{store} function will check the user-provide string \emph{buf} against predefined constants (Lines 10-12). If \emph{buf} passes the validation, the corresponding attribute will be updated, and therefore affect the driver code execution (Lines 23-27).

In sum, besides system calls and hardware interrupts, drivers' execution can also be affected by two types of variables: kernel module parameters and device attributes. It is worth pointing out that \emph{a driver is not only affected by the kernel module parameters to which it belongs and the device attributes of the bound devices but also by the device attributes of other adjacent devices}.

\section{Motivation}

This section provides a comprehensive study of runtime parameters. \emph{Firstly}, we conduct an experiment to quantify the number of branches affected by device attributes and kernel module parameters. We compare these with the code influenced by the system call arguments, and conclude that runtime parameters have a significant impact on control flow; \emph{Secondly}, we demonstrate a historical bug as our motivation example and provide our insights on how to incorporate these parameters into kernel fuzzing.

\subsection{The Impact of Runtime Parameters}

We are curious about how many codes are influenced by runtime parameters. We choose to count the number of affected branch statements (if/switch statements) and basic blocks, for the reason that branch coverage and basic block coverage are the most prevalent indicators used in fuzzing. To make the results more persuasive, we also count the branch that could be affected by syscall arguments, which can also be directly controlled by the user. The experiment details are listed in \autoref{exp_setup}, and the results are listed in \autoref{comparison}.

As demonstrated by the results, device attributes and kernel module parameters influence a considerable number of \texttt{if}/\texttt{switch} statements compared to data passed through syscall arguments, potentially contributing to code coverage during fuzzing. Interestingly, the number of branch statements users can control via syscall interfaces is surprisingly lower than anticipated. We conclude this for three primary reasons: 


\begin{packeditemize}

\item \textbf{R-1}: The isolation between kernel space and user space is a fundamental design principle to restrict user programs' access to kernel space to mitigate security risks. 

\item \textbf{R-2}: Kernel functionalities, such as task scheduling and memory management, are inherently uncontrollable to user programs. However, the code coverage of these functionalities is still counted during fuzzing, leading to higher coverage than the user-provided data can control.

\item \textbf{R-3}: The number of affected branch statements and basic blocks do not fully represent the extent of code coverage, as function calls within basic blocks may further expand the affected code range. Due to the limitation of static analysis, we drop the callee functions inside basic blocks to avoid imprecise indirect call analysis, and only count the directly affected branch and basic blocks for all three kinds of user data.
\end{packeditemize}

\begin{table}[!t]
\centering
\caption{The number of the branches and basic blocks affected by different user-controlled data.}
\resizebox{0.95\columnwidth}{!}{
\begin{tabular}{lllll} 
\toprule
 \textbf{Taint Source} &  \textbf{Number} & \textbf{Aff. If} & \textbf{Aff. Switch} & \textbf{Aff. BB}   \\
          
\hline
 Syscall Args        & 12839  & 30588 & 1036 & 74172 \\
 Module Params       & 1494   & 15790 & 46   & 35178 \\
 Device Attrs        & 1407   & 9226  & 58   & 20875 \\
\hline
\end{tabular}
}
\label{comparison}
\end{table}

To this point, we conclude that there is a large amount of code that remains uncovered due to runtime parameters. Although theoretically, these parameters can be modified by existing syscall interfaces such as open() and write(), this approach is impractical for current fuzzers due to the complexity of generating valid strings for file paths. Moreover, even if files under \emph{sysfs} are accidentally opened during testing, writing random values to them is unlikely to yield meaningful results, as the parameters are governed by value-checking logic (as shown in \autoref{figure4}). As of this writing, only 39 files have been explicitly included in Syzkaller, accounting for less than 2\% of the total runtime parameters.

\subsection{Motivated Example}
We will first show a historical bug related to runtime parameters, analyzing the root cause of it and the prerequisite of triggering it. Based on this example, we summarize our insights on integrating these parameters into kernel fuzzing.

\begin{figure}
\centering
\includegraphics[width=\columnwidth]{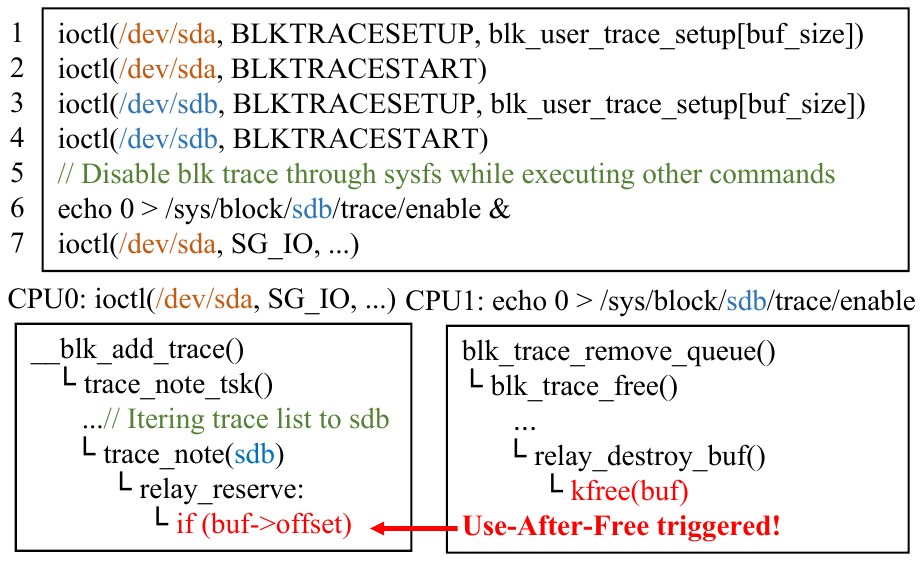}
\caption{PoC and call stacks related to CVE-2021-47375. The upper half is the proof-of-concept to trigger the bug, while the lower half depicts the root cause of it.}
\label{CVE-2021-47375}
\Description{Historical bug that unveiled the relation between devices.}
\end{figure}

\bheading{CVE-2021-47375.}
A previously discovered bug relating to runtime parameters is CVE-2021-47375, illustrated in \autoref{CVE-2021-47375}. The top portion of the figure presents the proof-of-concept code for this vulnerability. Lines 1-4 enable the BLKTRACE functionality for both hard disk devices, \emph{sda} and \emph{sdb}, causing the kernel to create corresponding nodes in a global list. Subsequently, on Line 6, the BLKTRACE functionality for \emph{sdb} is disabled via \texttt{sysfs}, while commands are simultaneously sent to \emph{sda} using \texttt{ioctl}. 

The bottom portion of \autoref{CVE-2021-47375} depicts the call stacks when triggering the bug. On CPU0, the \texttt{ioctl()} invokes \texttt{trace\_note\_tsk()}, which iterates over all nodes in the global trace list. Concurrently, CPU1 executes \texttt{blk\_trace\_remove\_queue()}, removing \emph{sdb} from the trace list. However, CPU0 remains unaware of this removal and attempts to access the \texttt{buf$\rightarrow$offset} field of the now-deleted node, resulting in a Use-After-Free vulnerability.

This example inspires us to believe that runtime parameters will not only affect the current device driver it is attached to, but also drivers of all devices with underlying data structures interconnected. In this case, changing device attributes for \emph{sdb} will result in crashing on executing commands related to \emph{sda}. Also, parameter modification alone cannot trigger the vulnerability. It should combine with the driver execution, and the time of modification is important for triggering the bug.

\bheading{Conclusion.}
Based on the statistics above and the motivated example, we conclude that runtime parameters are critical in kernel fuzzing. Existing fuzzing tools have not effectively utilized these interfaces, failing to uncover latent bugs. By analyzing historical vulnerabilities, incorporating runtime parameters with syscall invoking drivers will expose more vulnerabilities. We highlight that the relationships between devices, and the timing of modifications are essential factors in triggering kernel bugs.

\section{Design}
As mentioned, to test the kernel driver in-depth, we need to integrate runtime parameters, as well as the relationship among them, into fuzzing. However, incorporating the runtime parameters into the fuzz engine is non-trivial. It still faces three technical challenges, as listed below respectively: 

\begin{packeditemize}
    
\item \textbf{C-1:} How to identify runtime parameters and their corresponding values to pass input validation? Current interfaces can open and write files under the filesystem, but the kernel has strict check logic to prevent arbitrary values from being written to parameters. Randomly generated values may fail to modify the parameter, thereby hindering the execution of deep driver code.
    
\item \textbf{C-2:} How to determine relations between runtime parameters and drivers? The kernel encompasses thousands of associated parameters, and there will be exponential combinations if there is no guidance, decreasing fuzzing efficiency.

\item \textbf{C-3:} How to integrate the runtime parameters into the fuzz engine and cooperate with existing features? Setting runtime parameters will affect not only the current test case but also the following cases until the system reboots. This will result in unreproducible coverages for the following test cases, affecting other components in the fuzz engine.
\end{packeditemize}

In the following subsections, we will first propose an overview of our design, and explain how the design addresses the above three challenges. Then we will give out the details of each component, which resolves the challenges respectively.

\begin{figure*}[!t]
\centering
\includegraphics[width=\textwidth]{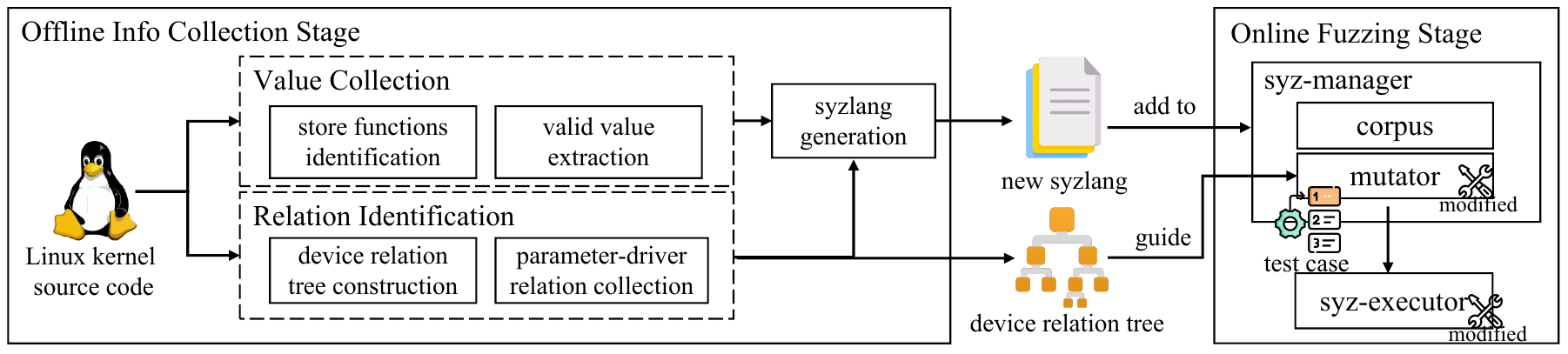}
\caption{The framework overview of \sys.} 
\label{figure6}
\Description{Framework overview of our work, including offline info collection and online fuzzing stages. The first stage collects the necessary info to integrate parameters into fuzzing, while the second stage utilizes it to fuzz.}
\end{figure*}

\subsection{Overview}\label{e4.1}

To address the aforementioned challenges, we propose a new fuzzing framework. The overall framework of \sys is depicted in \autoref{figure6}, comprising two primary stages: the \emph{offline information collection stage} and the \emph{online fuzzing stage}. 

In the first stage, \sys conducts the \textbf{value collection task} to overcome the \textbf{C-1}, collecting the filename and valid values from the source code to pass the input validation, thus modifying parameters successfully during the next stage. These values are subsequently used in \emph{syzlang} generation. To resolve the \textbf{C-2}, \sys performs the \textbf{relation identification task} to identify the two kinds of relationships, \emph{inter-device} and \emph{parameter-driver} relationships, and constructs a device relation tree. These relations will be used to guide mutators and narrow the search space.

In the second stage, \sys conducts the \textbf{fuzz engine integration task} to incorporate runtime parameters into fuzzing. We extend the fuzz engine and leverage relationships to prioritize the newly generated \emph{syzlangs}, facilitating more in-depth testing of relevant drivers. By utilizing existing interfaces in Syzkaller, we propose new mutation strategies that determine the parameter before driver execution, therefore solving the \textbf{C-3}. 

\subsection{Value Collection}\label{e4.2}

We need the filenames and valid inputs to change the value of the runtime parameters successfully. The filenames are predefined strings that are registered in global variables in kernel source code. By identifying the definition points, we can recover the filename from the variable. To obtain valid values for parameters, \sys first identifies store functions responsible for updating the parameter values, then extracts valid values according to the validation logic embedded within the store functions.

As demonstrated in previous sections, the \emph{struct device\_attribute} encapsulates the device attribute's name (which is also the filename in \emph{sysfs}) and associated \emph{store} function. Based on the design pattern of runtime parameters, we degrade the value collection task into two steps: store function identification and valid values extraction.

\subsubsection{Store function identification} 
The key to gathering filename-function pair is to identify globally defined \emph{struct device\_attribute} variables. Although in most cases these variables are initialized by the \texttt{\_\_ATTR()} macro provided by the Linux kernel, some device drivers define their device attribute structures. Since parameters can only be exported to the user space through \emph{struct attribute}, we manually inspect all the kernel structures that contain \emph{struct attribute} in kernel source code (56 results in total) and summarize the characteristics of all kinds of attribute variables:
\begin{packeditemize}
\item First, they are defined in the kernel as a series of global variables initialized when the driver is loaded.
\item Second, the show or store functions must be provided. Otherwise, it cannot be accessed through the file system. 
\end{packeditemize} 
For those driver-specific device attribute structures, at least one subfield matches the above rules. Thus, we iterate all global variables in the kernel and gather structure type variables. Following the above rules, we check every subfield recursively. Since read-only parameters will have limited influence on drivers' control flow, we omit all device attributes without writable permission.

\begin{algorithm}[!t]
\caption{Store Functions Identification}
\label{alg:algorithm1}

\begin{algorithmic}[1]
    \Require LLVM bitcode of Linux kernel
    \Ensure $store\_functions$, the list of pairs between store functions and attribute names
    \Function{matchDeviceAttr}{$var$}
        \If{$hasStructAttr($var$) \And hasStoreFuncPtr($var$)$}
            \State \Return $True$
        \EndIf
        \State \Return $False$
    \EndFunction
    \Procedure{getStoreFunc}{$var$}
        \If{$isStructTy(var)$}
            \If{\Call{matchDeviceAttr}{$var$}}
                \State $pair \gets getNameFuncPair(var)$
                \State $store\_functions.push(pair)$
            \Else
                \For{each, $elem \in var$}
                    \State $struct\_elem \gets getInitializer(elem)$
                    \State \Call{getStoreFunc}{$struct\_elem$}
                \EndFor
            \EndIf
        \EndIf
    \EndProcedure
    \For{each, $gv \in GlobalVariables$}
        \State \Call{getStoreFunc}{$gv$}
    \EndFor

\end{algorithmic}

\end{algorithm}

\bheading{Algorithm.} Building on the understanding mentioned above, we propose the following algorithm, with the corresponding pseudo-code presented in Algorithm~\ref{alg:algorithm1}. The \texttt{for} loop in Lines 20-22 iterates over all global variables for analysis, calling the procedure ``GetStoreFunc()'' to verify each variable. In the procedure, Line 8 verifies whether the input variable is structure type. Subsequently, the \emph{matchDeviceAttr()} function is called to determine whether the variable includes the attribute structure along with pointers to the associated \emph{show} and \emph{store} functions (Lines 1–6). If this condition is satisfied, the algorithm traverses all elements of the variable to locate the \emph{store} function, pairs it with the attribute's name, and appends the result to the list \emph{store\_functions} (Lines 9–11).  It is important to note that the file name, the device attribute's name, and the variable's name are not always identical. Therefore, we specifically extract the file name string as defined within the \emph{struct device\_attribute}. For structures that do not meet the criteria, the algorithm recursively iterates through all sub-fields to identify the target structure, as depicted in Lines 13–15.

\subsubsection{Valid Values Extraction.}

We construct a static taint analysis pass targeting all the identified \emph{store} functions to extract the variable types and the corresponding range of valid values. By marking the user-provided \texttt{buf} as the taint source and analyzing the sink points along the control flow, we identify the functions responsible for validation. For string type parameters, the check logic resides in string comparison functions (e.g., \texttt{match\_string()}). For parameters of other types, the kernel invokes helper functions to convert the original string(e.g., \texttt{kstrtouint()}). In such cases, we track the converted value and its sink points instead of the original user-provided string. Subsequently, we determine the range of values that can successfully pass the validation logic.

Since kernel module parameters do not have input validation, we collect all parameters and their relative types by identifying their definition points. Linux uses the macro \texttt{module\_param()} to define kernel module parameters. The macro will finally be expanded to a structure containing parameter names and types, which are sufficient for us to construct \emph{syzlang}.

\subsection{Relation Identification}\label{e4.3}

After identifying the parameter file names, their corresponding store functions, and their valid values, the next objective is to build up the relationship between parameters and related drivers. Based on our understanding of parameter-related vulnerabilities, the relationship contains two perspectives: \emph{Inter-device relationship.} Since devices are organized by device tree through underlying data structures, the device's attributes will affect topologically connected devices; \emph{Parameter-driver relationship.} Modification of the parameters alone is insufficient to trigger bugs. It should combine with related driver code execution. The following sections will introduce our strategies for collecting relations mentioned above.

\subsubsection{Device Relation Tree Construction.}
The \emph{sysfs} interface exports devices in a hierarchical structure, enabling the construction of a device relation tree through folder traversal. Specifically, a sub-device appears as a sub-folder within the parent device's directory in the \emph{sysfs}, with the root directory corresponding to the bus to which the device is attached. The device relation tree is constructed with the bus as the root node.

\subsubsection{Parameter-Driver Relation Collection.}
To trigger the execution of related drivers after modifying parameters, it is necessary to establish a mapping between parameter files and their corresponding drivers. In Linux, parameter files reside in the device's folder under \emph{/sys}, and the user can invoke drive execution by opening and operating files under the \emph{/dev} directory. Consequently, for all devices located in the \emph{/dev} directory, the relationship between their parameters and the drivers they bond to can be established by matching their file names with the directory names under \emph{/sys}.

Based on the files and values collected in the previous steps, we adopt the methodology outlined in existing work~\cite{hao2023syzdescribe} to generate \emph{syzlangs}, facilitating the modification of runtime parameters. To minimize redundancy caused by multiple devices of the same type (e.g., \texttt{usb0} and \texttt{usb1} sharing identical parameter types), we merge parameters of the same type into one \emph{syzlang}.

\subsection{Fuzz Engine Integration}\label{e4.4}

We should not only enable the parameter modification but also make sure the relations collected above are integrated into the fuzz engine. Therefore, two primary components have to be changed accordingly: The \emph{executor} should have extra primitives that prepare the parameter before related driver execution; The mutator should utilize the relations to combine related parameters and schedule the execution timing to trigger more vulnerabilities. In the following sections, we will demonstrate our design for each component.

\subsubsection{Execution Engine Modification.}
To establish the association between drivers and parameters within Syzkaller, we employ the pseudo-syscall mechanism, which integrates parameter modifications with syscalls that invoke the corresponding drivers. As discussed above, pseudo-syscalls' design adheres to two key principles: First, pseudo-syscalls must be universal, as a single driver may have numerous runtime parameters. Creating unique pseudo-syscalls for each parameter would demand substantial computational resources to record and update \emph{syzlang} relations. Second, pseudo-syscalls should establish dependencies with other \emph{syzlangs} associated with the same driver to maintain coherence. We implement pseudo-syscalls within the execution engine and register them alongside existing syscalls to ensure seamless integration and functionality.

\subsubsection{Mutation Strategy Integration.}
As demonstrated in the motivating example (\autoref{CVE-2021-47375}), both the relationships between devices and the timing of parameter modifications are critical for uncovering kernel bugs. Regarding the combination of parameters, we prioritize the modification of parameters of topologically related devices at the same time. Concerning execution timing, we initiate asynchronous threads that modify parameters while executing other system calls targeting the corresponding device driver.

To this end, we incorporate fuzzing with the runtime parameters, exposing long-neglected code functionalities. By integrating new \emph{syzlangs} and proposing new mutation strategies, we extend Syzkaller's ability to test kernel drivers.

\section{Implementation}

We implemented a prototype of \sys, and the overall lines of code of our implementation are presented in \autoref{LoC}. As shown in the table, the static analysis of kernel global variables and functions is based on LLVM\cite{LLVM} and CRIX\cite{lu2019detecting}. The implementation of syzlang generation based on SyzDescribe\cite{hao2023syzdescribe}. The incorporation of the fuzzing system is based on Syzkaller\cite{Syzkaller}. We will illustrate the implementation details of each component below.

\begin{table}[!t]
\centering
\caption{The statistics of the lines-of-code of each component.}
\resizebox{\columnwidth}{!}{
\begin{tabular}{llll} 
\toprule
 \textbf{Component} &  \textbf{LoC} & \textbf{Language}  & \textbf{Implemanted on} \\
          
\hline
 Attribute value collection           & 1598  & C++ & LLVM, CRIX  \\
 Syzlang generation                   & 396   & C++ & SyzDescribe \\
 Fuzz engine implementation           & 252   & C / golang & Syzkaller   \\
 Other Scripts                        & 391   & C / Codeql & -        \\
\hline
\end{tabular}}
\label{LoC}
\end{table}

\bheading{Value Collection.}
We develop a static analysis tool to examine kernel source code and extract predefined values. Using customized LLVM module passes, we track definition points of global variables of \emph{struct device\_attribute} type. For driver-defined structures, we recursively traverse sub-fields within them according to the algorithm. Then, we locate the attribute value verification logic within the corresponding attribute store functions. For cases involving indirect calls in the store functions, we utilize CRIX's type-based function indirect call analysis tool to resolve the target functions of these calls. By performing a static taint analysis and matching sink points of user-provided strings, we determine whether the validation involves string comparisons or system string conversion utilities. We collect string comparison targets as valid values in the first case, and record the variable type after conversion otherwise.

\bheading{Relations Identification.}
For the device relation tree construction, the \texttt{parent} field of \emph{struct device} is assigned when a device is registered to the system, and device files are only created during system startup. Consequently, these relationships can not be determined using static analysis. To address this, we develop scripts to traverse the \emph{/sys} directory within the virtual machine, capturing all device directories and their hierarchical relationships using a depth-first approach. During this traversal, we catalog all parameter files in the device directories and map them to the previously identified structures. This process comprehensively maps device files and their corresponding value ranges.

For parameter-driver relationship identification, we collect device files within the \emph{/dev} directory and associate them with their corresponding device folders in the \emph{sysfs} based on matching file names. This approach enables an accurate mapping between parameter files and the associated drivers.

\bheading{Syzlang Generation.}
We construct new syzlangs for each parameter type in this module. Due to different device instances sharing the same parameter type, we generate one syzlang for all relevant instances that have the parameter type to minimize the number of newly generated syzlangs. For kernel module parameters, we iterate over all files under \emph{/sys/modules} directory and match them with parameters collected before to determine their types.

\bheading{Execution Engine Modification.} 
We utilize the pseudo-syscall interface provided by Syzkaller to obtain maximum compatibility. Our pseudo-syscall is defined as \emph{syz\_mod\_dev(param\_path, param\_val, dev\_path, intptr, flags)}. Arguments include the path to the parameter file, the modified value for the target parameter, the path to the device file, a data pointer used as a random seed, and the open permissions for the device file. By leveraging the established mapping relationship between the \emph{/dev} and \emph{/sys} directories, we generate parameter combinations for the pseudo-syscall using the associated attribute and device files. 

We incorporate the C functions representing newly defined pseudo-syscalls into Syzkaller's executor module. The main functionality is modifying parameters, opening the associated device file in the \emph{/dev} directory, and returning its file descriptor. Notably, many buses initialize device names using a naming convention such as "bus name - device number". To address the issue of generating homogeneous syzlangs for each device, we utilize the wildcard ``\texttt{\#}'' to represent any device within the same bus, selecting a random device during fuzzing. The data pointer in arguments is used as a random source. To ensure consistency, the same serial number is employed to select both the target device and the corresponding parameter files. This approach maintains alignment between the files while optimizing fuzzing efficiency.

\bheading{Mutation Integration.}
We enhance Syzkaller's existing mutator by introducing a novel mutation strategy. Specifically, when mutating a seed containing newly defined syzlangs, we integrate a complementary syzlang that modifies the parameters of the devices on the same device tree. These syzlangs are executed concurrently on different threads during driver executions, trying to trigger the race condition between driver execution and parameter modification of related devices. This approach facilitates a comprehensive exploration of device interactions during fuzzing.
\section{Evaluation}

In this section, we evaluate \sys by comparing it with related approaches and presenting comprehensive experiment results. Specifically, we aim to address the following questions:
\begin{packeditemize}
    \item How many device attributes can be extracted during the offline information collection stage? (\autoref{e6.2})
    \item Can the incorporation of runtime parameters enhance the code coverage of kernel drivers during fuzzing? (\autoref{e6.3})
    \item Does leverage relationships between devices contribute to the identification of vulnerabilities? (\autoref{e6.4})
    \item Can new vulnerabilities, previously undetectable from the perspective of the driver model, be identified? (\autoref{e6.5})
\end{packeditemize}

\subsection{Experiment Setup}\label{e6.1}

All experiments were conducted on a server equipped with two Intel Xeon Gold 6430 CPUs and 256GB of RAM, running Ubuntu 22.04 LTS as the platform system. The target Linux kernel version was v6.7-rc7, released in January 2024. Our fuzzing framework was developed based on Syzkaller\cite{Syzkaller} (commit c52bcb237b7f), with Clang-14 utilized as the compiler to generate LLVM IR. Both the offline and online stages employed the same Linux kernel configuration, which is also used by the upstream kernel in Syzbot. These kernel compilation options, provided by Google engineers, reflect their understanding of testing the Linux kernel with QEMU. The kernel configuration option \texttt{CONFIG\_SYSFS} is enabled by default. To ensure comprehensive coverage, we manually enabled all \texttt{sysfs}-related configuration options (ten in total).

For comparative evaluation, we selected state-of-the-art kernel fuzzing tools, including Syzgen++\cite{chen2024syzgen++} and Syzdescribe\cite{hao2023syzdescribe}. Syzgen++ extracts syzlangs from kernel image files using symbolic execution and dynamic debugging techniques, aiming to improve code coverage by analyzing relationships between syzlangs. Syzdescribe applies static analysis techniques to identify system call-related code in the kernel source, gather critical information, and generate new syzlangs for fuzzing. Since Syzgen++ was initially tested on the Linux kernel version 5.15, we adapted it for compatibility with the target kernel version 6.7.

\begin{table*}[!ht]
\centering
\caption{The edge coverage results for selected drivers.}
\resizebox{\textwidth}{!}{
\begin{tabular}{l|ccccc|ccc} 
\toprule

\multirow{2}{*}{\textbf{Drivers}} & \multicolumn{5}{c|}{\textbf{\#Avg. Edge Coverage}} & \multicolumn{3}{c}{\textbf{p-value}}  \\
 & \textbf{Syzkaller} & \textbf{SyzDescribe} & \textbf{SyzGen++} & \textbf{\sys-} & \textbf{$\uparrow$Comparing to Syzkaller} & \textbf{Syzkaller} & \textbf{SyzDescribe} & \textbf{SyzGen++} \\
\hline
rtc     & 3,424  & 3,537  & 3,517  & \textbf{5,337}  & 55.89\% & 0.0060 & 0.0040 & 0.0040   \\
usb     & 11,301 & N/A   & 11,638 & \textbf{16,523} & 46.21\% & 0.0040 & N/A    & 0.0040    \\
loop    & 6,348  & 7,708  & 7,120  & \textbf{9,244}  & 45.63\% & 0.0040 & 0.016  & 0.0040   \\
msr     & 1,883  & 2,100  & 2,474  & \textbf{2,529}  & 34.28\% & 0.0040 & 0.0040 & 0.0040   \\
i2c     & 3,481  & 3,297  & 4,309  & \textbf{4,386}  & 26.02\% & 0.0040 & 0.0040 & 0.21   \\
sr      & 7,301  & 6,954  & N/A   & \textbf{8,653}  & 18.52\% & 0.0040 & 0.0040 & N/A   \\
mouse   & 2,958  & 3,690  & N/A   & \textbf{3,482}  & 17.73\% & 0.0040 & 0.0040 & N/A  \\
seq     & 3,610  & 3,690  & \textbf{4,244}  & 4,198  & 16.29\% & 0.0040 & 0.0278 & 0.85  \\

\hline
\end{tabular}}
\label{table3}
\end{table*}

\subsection{Results of Static Analysis}\label{e6.2}

Our tool analyzed all device attribute variables in the target kernel version, collecting their respective store functions and valid input. In total, 1,243 device attributes were identified. Among these, 358 attributes accept unsigned integers, 229 accept signed integers, 107 accept boolean values, 60 accept strings, and 80 accept formatted string inputs. Note that the original inputs provided by users are always strings, and the types listed above stand for the value after type conversion during input validations. Additionally, 46 device attributes were found to execute disregarding user-provided values. However, the input legality could not be determined for 363 functions due to the following primary reasons:

\begin{packeditemize} 
    \item \textbf{Value propagation and indirect calls.} Store functions propagate the \emph{buf} value to other functions via indirect calls. Due to the inherent limitations of static analysis, identifying the target functions of these indirect calls was not feasible. 
    \item \textbf{Complex parameter processing logic.} Some input validation logic processes the user-provided string byte-by-byte using loops. This intricate logic currently exceeds the analytical capabilities of our method, preventing the determination of feasible values for these device attributes. 
\end{packeditemize}

As for kernel module parameters, our tool finds 694 unique parameters, including 26 string-type parameters. The number of runtime parameters is less than the data collected in previous sections, with the reason we use fuzzing config here instead of the all-yes-config used previously. Also, the all-yes-config will include runtime parameters for different architectures, which will never appear in a single test run.

We then gathered the mappings of all files and directories within the \emph{/sys} directory in the VM at the offline profiling stage and found a total of 17,081 writable files. We extracted the filenames of these attribute files and matched them with the name field in the collected device attribute structures. Subsequently, we associated these files with their respective device attribute store functions, recording 819 types of files that have their corresponding functions. We noticed that 424 store functions that do not pair with files under the \emph{/sys} directory. This is because although the driver defined the parameters interface, there is no corresponding device bond to the driver, thus failing to create parameter file instances.

\bheading{Parameter name conflict.} Since different types of devices may share the same filename for different device attributes, we iterate through all device attributes, identifying 55 instances where attributes share the same name. Among these, 41 pairs exhibited identical attribute types. However, we did find 14 pairs of attributes that have different attribute types. Since it only accounts for 1.71\% of the total number of device attributes, we argue that the conflict caused by filename collision has minor effects on the result.

For runtime parameters that cannot be distinguished, our current approach uses random strings to populate the buf parameter when generating syzlang. To enhance the accuracy of our analysis in the future, we plan to integrate techniques such as symbolic execution to gather more precise values of device attributes.

\subsection{The Code Coverage}\label{e6.3}

In this subsection, we compare our work with other related work to demonstrate how incorporating runtime parameters can enhance code coverage during kernel driver fuzzing. We selected eight drivers on different buses from the list of drivers supported by Syzgen++ and Syzdescribe. These drivers also supported runtime parameters that can be modified. 

\bheading{Experiment setup.} To ensure fairness, we modified our tools to produce only syzlang for tuning parameters that the original Syzkaller can directly parse. Two primary considerations drove this decision: 1) All other related works in the experiments utilized the unmodified Syzkaller; 2) Our modified mutator established relationships between devices, and it would introduce coverage from other device drivers, which is not fair for comparison. Hence, we provided syzlangs generated by our tool(marked as \sys-), Syzgen++, and Syzdescribe to the same version of Syzkaller, and turned on only the syzlang of the corresponding driver. During the test environment setup, we identified some errors in the device file names and paths within the syzlangs generated by Syzdescribe. To ensure fairness, we manually corrected these errors. For each driver, we deployed 2 VMs with two vCPUs per VM, executing each test for 24 hours, totaling five runs. The number of syzlang, device attributes, and kernel module parameters enabled for each driver are listed in \autoref{param_num} in the appendix. The experiment results and corresponding confidence p-values are presented in \autoref{table3}, in which the code coverage stands for edge coverage.

\begin{table}[h!]
\centering
\caption{Comparison for \sys, Syzgen++ and combined.}
\resizebox{\columnwidth}{!}{
\begin{tabular}{lcccc} 
\toprule

\textbf{Drivers}             & \textbf{Syzgen++} & \textbf{\sys}- & \textbf{Combined} & \textbf{$\uparrow$ in Edge Cov.}  \\
\hline
rtc               &    3,517     &    5,337    &   5,560   &   58.09\% \\
usb               &    11,638    &    16,523   &   17,671  &   51.84\% \\
loop              &    7,120     &    9,244    &   9,227   &   29.59\%   \\
msr               &    2,474     &    2,529    &   3,158   &   27.65\%    \\
i2c               &    4,309     &    4,386    &   4,742   &   10.05\%    \\
seq               &    4,244     &    4,198    &   4,759   &   12.13\%   \\
\hline
\end{tabular}}
\label{table4}
\end{table}

\bheading{Edge coverage results.} The results show that \sys achieves the highest edge coverage among all related works in 7 out of 8 drivers. \sys outcompetes original Syzkaller in all 8 drivers and obtains an average coverage improvement of 32.57\%. 
According to the Mann-Whitney U Test, the results for most drivers are significant, with a p-value lower than 0.05.

However for certain drivers, such as \emph{i2c} and \emph{seq}, the p-value exceeds the threshold when compared to SyzGen++. To further validate that our coverage improvements are not merely a subset of those achieved in previous research, we conducted an additional experiment combining our syzlangs with those of SyzGen++, with the same setup. The edge coverage results, presented in \autoref{table4}, demonstrate that the combined syzlangs outperform SyzGen++ in terms of edge coverage across all six drivers supported by both approaches. These findings indicate that our method not only enhances the performance of existing techniques but also achieves higher coverage by exploring new code paths rather than retesting the same portions of the codebase.

Given the result that the coverage improvement is not evenly distributed among drivers, we conclude that this is because the functionalities of different buses vary significantly, leading to variations in the number and types of parameters they possess. For block device drivers like \texttt{loop}, the runtime parameters can control the IO status including poll rate and sector bytes. However, in drivers such as \texttt{i2c}, runtime parameters only have limited control, including power management and device notification.

\subsection{The Number of Vulnerabilities}\label{e6.4}

\begin{figure}
\centering
\includegraphics[width=\columnwidth]{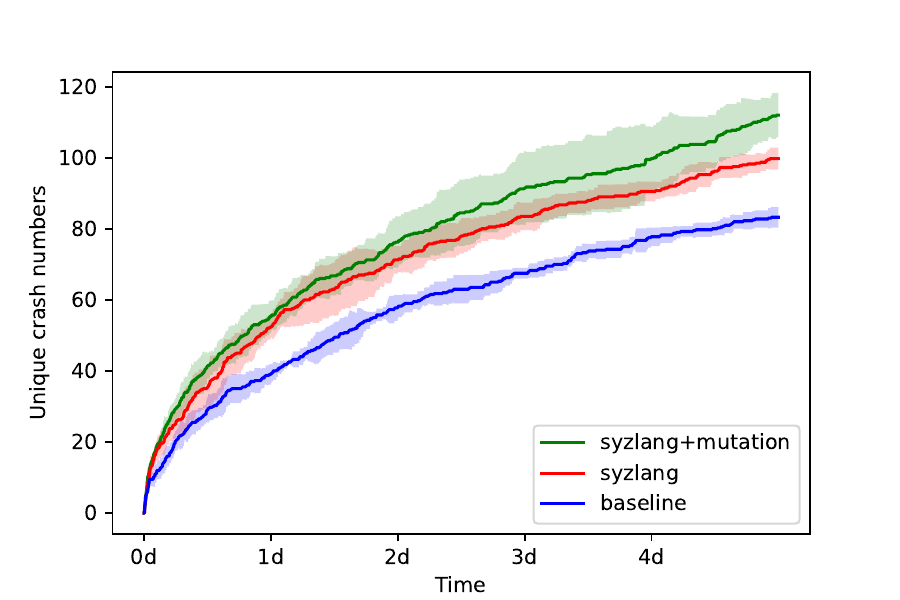}
\caption{Number of unique crashes over time.}
\label{figure7}
\Description{Number of unique crashes over time, which demonstrate the effectiveness of syzlangs we provide and mutation strategies we proposed.}
\end{figure}

To illustrate the contribution of the syzlang we generate and the mutation strategy we proposed to vulnerability discovery, we did an ablation study on these components. We choose the number of unique crashes as the indicator of vulnerability discovery ability. The overall results are shown in \autoref{figure7}. The original Syzkaller is used as baseline (marked as ``baseline''), compared to a version that adds syzlangs alone (marked as ``syzlang'') and another version that combines syzlangs with modifications to the fuzzing mutation strategy (marked as "syzlang+mutation"). Each fuzzer uses 8 QEMU VMs with 2 vCPUs per VM, conducting an experiment lasting 120 hours (5 $\times$ 24h), totaling four rounds of experiments. We count the number of UNIQUE crashes as well as the discovery time and calculate with the confidential interval of 0.90. Unique crashes are distinguished by their crash title, including their crash types and the function that triggers them. To mitigate any impact from Syzkaller's inherent issues, we ignore crash types beginning with ``SYZFATAL''.

The results demonstrate that the incorporation of syzlang for modifying runtime parameters facilitates the discovery of previously unknown vulnerabilities. Additionally, leveraging device relationships to guide fuzzing mutations enhances the efficiency of vulnerability discovery. Over 120 hours, the two optimizations that we propose improve the number of vulnerabilities found significantly. Adding syzlang alone increases the number of unique crash types by 19.8\% compared to baseline, while adding syzlang and mutation strategy gains 34.5\% improvement in average.

\begin{table*}[!t]
\centering
\caption{Previous-unknown bugs found by our tool.}
\begin{threeparttable}
\resizebox{\textwidth}{!}{
\begin{tabular}{rllllc} 
\toprule

 \textbf{ID} & \textbf{Bug Type} & \textbf{Error location} & \textbf{Subsystem} & \textbf{Status}\tnote{1} & \textbf{Related to parameters?} \\
\hline
\href{https://lore.kernel.org/linux-cve-announce/2024030409-CVE-2024-26622-9e01@gregkh/}{\#1}      & KASAN slab out-of-bound / UAF   & tomoyo\_write\_control             & security/tomoyo                      & CVE-2024-26622                & \ding{51}   \\
\href{https://lore.kernel.org/linux-cve-announce/2024050123-CVE-2024-26933-c18d@gregkh/}{\#2}      & task hung (deadlock)            & hub\_activate                      & drivers/usb/core                     & CVE-2024-26933/26934\tnote{2} & \ding{51}   \\
\href{https://lore.kernel.org/linux-cve-announce/2024050115-CVE-2024-27055-449e@gregkh/}{\#3}      & divide-by-zero                  & wq\_update\_node\_max\_active      & kernel/workqueue                     & CVE-2024-27055                & \ding{51}   \\
\href{https://lore.kernel.org/linux-cve-announce/2024053034-CVE-2024-36896-783f@gregkh/}{\#4}      & nullptr deref                   & disable\_store                     & drivers/usb/core                     & CVE-2024-36896                & \ding{51} \\
\href{https://lore.kernel.org/all/2024062137-CVE-2024-37356-cc7b@gregkh/}{\#5}      & UBSAN: shift-out-of-bounds      & dctcp\_update\_alpha               & drivers/net/ipv4                     & CVE-2024-37356                & \ding{51} \\
\href{https://lore.kernel.org/linux-cve-announce/2024062035-CVE-2024-38619-97c7@gregkh/}{\#6}      & divide zero                     & alauda\_transport                  & drivers/usb/storage                  & CVE-2024-38619                & \ding{55} \\
\href{https://lore.kernel.org/linux-cve-announce/2024070512-CVE-2024-39472-f977@gregkh/}{\#7}     & KASAN slab-out-of-bound         & xlog\_cksum                        & fs/xfs                               & CVE-2024-39472                & \ding{55} \\
\href{https://lore.kernel.org/linux-cve-announce/2024070912-CVE-2024-39487-f52c@gregkh/}{\#8}      & KASAN out-of-bounds read        & in4\_pton                          & drivers/net/bonding                  & CVE-2024-39487                & \ding{51} \\
\href{https://lore.kernel.org/linux-cve-announce/2024051943-CVE-2024-35876-d9b5@gregkh/}{\#9}      & ODEBUG\_STATE\_WARNING          & mcheck\_cpu\_init\_timer           & arch/x86/kernel/cpu                  & fixed                & \ding{51} \\
\href{https://lore.kernel.org/all/20240302064312.2358924-1-wangkefeng.wang@huawei.com/}{\#10}      & UBSAN: shift-out-of-bounds      & fault\_around\_bytes\_set          & mm                                   & fixed                & \ding{51} \\
\href{https://lore.kernel.org/all/20240305031221.492421-2-paul@paul-moore.com/}{\#11}     & warning                         & sel\_write\_load                   & security/selinux                     & fixed                & \ding{51} \\
\href{https://lore.kernel.org/all/CAEkJfYOt60_VqdcepU0pHa-NctFu8AvnWQHKZ+WAZtheAPBXzg@mail.gmail.com/}{\#12}     & UBSAN: array-index-out-of-bound & diFree                             & fs/jfs                               & fixed             & \ding{55} \\
\href{https://lore.kernel.org/all/20240610124406.422897838@linutronix.de/}{\#13}     & warning                         & static\_key\_slow\_inc\_cpuslocked & kernel/jump\_labal$+$arch/x86/events & fixed                     & \ding{51} \\
\href{https://lore.kernel.org/all/CAEkJfYMtSdM5HceNsXUDf5haghD5+o2e7Qv4OcuruL4tPg6OaQ@mail.gmail.com/}{\#14}     & KASAN: slab use after free read & zswap\_decompress                  & mm/zswap                             & fixed                & \ding{51} \\
\href{https://lore.kernel.org/all/CAEkJfYNguDt47=KnEUX7tLwx_46ggBx3Oh3-3dAcZxqndL_OWQ@mail.gmail.com/}{\#15}     & UBSAN: shift-out-of-bounds      & sg\_build\_indirect                & drivers/scsi                         & patch under revision & \ding{51} \\
\href{https://lore.kernel.org/all/CAEkJfYNYe5tSMfAn9K_zJ1O_Vu7jxJDZv_uYCgTW=NZiUzcAuw@mail.gmail.com/}{\#16}     & warning                         & ax25\_dev\_device\_down            & drivers/net/ax25                     & patch under revision & \ding{51} \\
\href{https://lore.kernel.org/all/CAEkJfYMMobwnoULvM8SyfGtbuaWzqfvZ_5BGjj0APv+=1rtkbA@mail.gmail.com/}{\#17}     & KASAN: slab use after free read & pressure\_write                    & kernel/cgroup                        & patch under revision & \ding{51} \\
\href{https://lore.kernel.org/all/CAEkJfYOyWgJW-WAd+GhT07zd2Y3vUWz81+pjbZT9nUAsCc7FGQ@mail.gmail.com/}{\#18}     & task hung (deadlock)            & rfkill\_fop\_write                 & net/rfkill                           & confirmed            & \ding{51} \\
\href{https://lore.kernel.org/all/CAEkJfYNNZftjpYBpnH4tEnm82orKtQ6SQn9i3sg7YNO-Df3tSQ@mail.gmail.com/}{\#19}     & warning on once                 & static\_key\_disable\_cpuslocked   & kernel/jump\_labal                   & confirmed            & \ding{51} \\
\href{https://lore.kernel.org/all/CAEkJfYPO8OK=JCFphuZvqzqCWpUjPiTVoHma3CY0gLo+rdLKNw@mail.gmail.com/}{\#20}     & task hung                       & blk\_mq\_get\_tag                  & drivers/block                        & confirmed            & \ding{55} \\
\href{https://lore.kernel.org/all/CAEkJfYMa7T7kudNECEhh_UOrVV1EpWqorWw5Pm2JxTRTsiaqFg@mail.gmail.com/}{\#21}     & task hung                       & lock\_sock\_nested                 & drivers/net                          & reported             & \ding{51} \\
\href{https://lore.kernel.org/all/CAEkJfYPLWixVyuKGoHHtPizvNN2nrfzwrxF71MXVZNgYmfmYDQ@mail.gmail.com/}{\#22}     & deadlock                        & ptp\_clock\_unregister             & drivers/ptp                          & reported             & \ding{51} \\
\href{https://lore.kernel.org/all/CAEkJfYMqyx4_13X5-SJYCJ5JnY6MbPR7Bi2mM_n7KG5Ng8-2fA@mail.gmail.com/}{\#23}     & warning                         & disksize\_store                    & drivers/block/zram                   & reported             & \ding{51} \\
\href{https://lore.kernel.org/all/CAEkJfYOA5C97jbzx-_UvGSqNE+ppjPmifyv45dJjKA5yx44Sgw@mail.gmail.com/}{\#24}     & warning                         & cfg80211\_bss\_update              & drivers/net/wireless                 & reported             & \ding{51} \\
\href{https://lore.kernel.org/all/CAEkJfYONxEew-9-KsXG_=C-Qx5j=27CB43JEPhiLaGrXenLHpA@mail.gmail.com/}{\#25}     & general protection fault        & udmabuf\_create                    & drivers/dma-buf                      & reported             & \ding{51} \\
\href{https://lore.kernel.org/all/CAEkJfYMCSkphb0Ax3mqBpOch6_BL0uEqAMNux1n=57fPChtGKQ@mail.gmail.com/}{\#26}     & KASAN: slab-out-of-bound        & asus\_report\_fixup                & drivers/hid                          & reported             & \ding{55} \\
\href{https://lore.kernel.org/all/CAEkJfYPXwvARuevbL-rWYWFs6NbrvbwvU7Hpf9xap6qpQjLcdw@mail.gmail.com/}{\#27}     & warning                         & read\_rindex\_entry                & fs/gfs2                              & reported             & \ding{55} \\
\href{https://lore.kernel.org/all/CAEkJfYN_F=NTByBdzjxqZJ7shcLjTcm4nXwX9GOhthDCAMPLSQ@mail.gmail.com/}{\#28}     & warning                         & ext4\_iomap\_begin                 & fs/ext4                              & reported             & \ding{55} \\
\href{https://lore.kernel.org/all/CAEkJfYM1nkT=xffLhYF-cCVKA3cPb2S80yqrBr=FsPPZRTcXOA@mail.gmail.com/}{\#29}     & kernel bug                      & folio\_end\_read                   & fs/gfs2                              & reported             & \ding{55} \\
\href{https://lore.kernel.org/all/CAEkJfYMMR4r1C2nWiZMOQKiVv8AQyApjgg-qt5UHghY0-w4sFQ@mail.gmail.com/}{\#30}     & general protection fault        & gfs2\_unstuffer\_folio             & fs/gfs2                              & reported             & \ding{55} \\
\hline
\end{tabular}}
\begin{tablenotes}
    \footnotesize
    \item[1] In the \textbf{Status} column, ``patch under revision'' means that the bug is confirmed and the patch is under review
    \item[2] This bug has multiple trigger places, fixed by two patches, each of them assigned a CVE number
    
\end{tablenotes}
\end{threeparttable}
\label{table5}
\end{table*}

\subsection{Bug Finding}\label{e6.5}

To demonstrate our tool's capability to discover new vulnerabilities, we conducted a long-term experiment. As shown in \autoref{table5}, we have identified a total of 30 previously undiscovered vulnerabilities, scattering in different subsystems. We reported these vulnerabilities to the Linux community, of which 20 have been confirmed, 14 have been fixed, and 9 CVEs have been assigned.

As shown in the table, 21 out of 30 bugs we found are related to runtime parameters. Most parameter-related bugs are located in subsystems that are closely related to the device, including \texttt{drivers}, \texttt{net}, \texttt{kernel}, etc. It is worth pointing out that, since \texttt{CPU} is a type of special device, it also appears under \emph{/sys/devices}, and the implementation is under \texttt{arch} subsystem. Among 21 bugs that are related to runtime parameters, 7 of them are exposed in concurrent behaviors, testifying our mutation strategy helps uncover bugs.

Moreover, introducing runtime parameters to fuzzing also helps us uncover various bug types, including use-after-free, out-of-bounds read, null pointer dereference, deadlock, etc. These bugs are security-related and can affect the kernel stability.

\subsection{Case Study}\label{casestudy}

In this subsection, we will use a specific vulnerability uncovered by \sys, CVE-2024-36896, to demonstrate its ability to identify vulnerabilities. This bug is caused by improper handling of inter-device relationships and should be triggered by concurrently accessing the runtime parameters.

\begin{figure}[!t]
\centering
\includegraphics[width=\columnwidth]{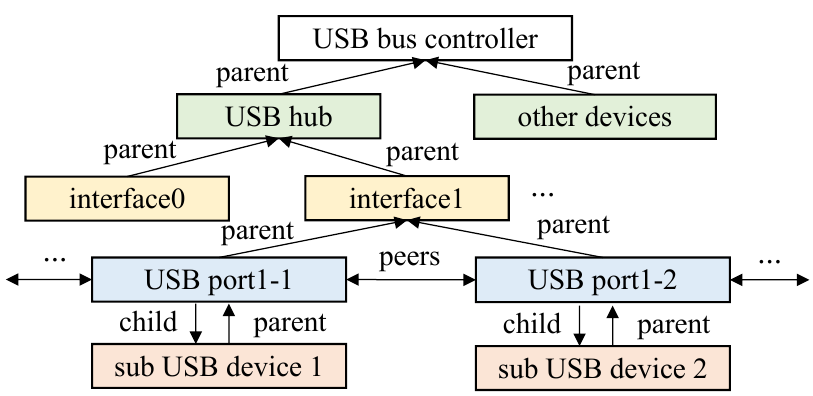}
\caption{Device relationships under USB bus.}
\label{figure8}
\Description{Relation between USB devices and bus.}
\end{figure}

\bheading{USB device hierarchy.} In kernel devices, USB devices are an excellent example to illustrate relationships between devices. USB devices can connect to the system through USB hubs, and a single USB device may have multiple functionalities (e.g., a webcam with both video and audio capabilities). The Linux kernel uses hierarchical structures to describe USB device organizations, such as \texttt{usb\_device}, \texttt{usb\_hub}, \texttt{usb\_port}, and \texttt{usb\_interface}. The USB driver uses \emph{parent} pointer to organize devices and form a device tree, with the USB bus controller as the root. The relationships among these device structures are illustrated in \autoref{figure8}.

\begin{figure}[!t]
\centering
\includegraphics[width=\columnwidth]{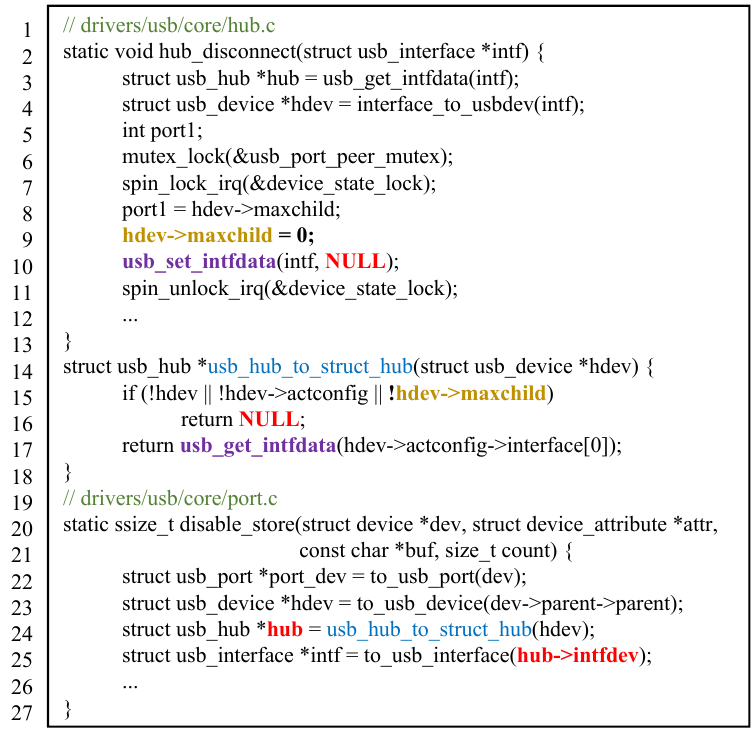}
\caption{Kernel source code of USB hub and USB port. The concurrent access triggers a null pointer dereference.}
\label{figure9}
\end{figure}

\bheading{Root cause analysis.} The kernel code related to the aforementioned vulnerability is displayed in \autoref{figure9}. When a USB hub device is removed from the system, the kernel invokes the \emph{hub\_disconnect} function (Line 2), which acquires the \texttt{device\_state\_lock} of the USB hub to prevent other devices from tampering with the data state (Line 7). The driver sets the maxchild number to zero and sets the data of \emph{struct usb\_interface} to NULL to avoid access (Lines 10-11). During removal, if someone attempts to access a subdevice's parameter under the hub (e.g., to access the \texttt{disable} parameter on one of the ports), the kernel invokes the store functions of that parameter. In this case, the store function \texttt{disable\_store()} will try to get the port device's parent hub, and get private data from the hub(Lines 22-24). \texttt{usb\_hub\_to\_struct\_hub()} checks whether the necessary data for the hub device exists, and returns the \emph{struct usb\_hub} pointer if passed the check. The \texttt{disable\_store()} function directly accesses the fields in the structure pointer without attempting to acquire the lock of the parent device or checking the return value, resulting in a null pointer dereference.

In this case, developers failed to take into account the relationship between devices. During the removal of the parent device, child devices can still access their parent device's data through \texttt{parent} pointer, without holding the parent's lock. Thus the code execution of the child device's driver is affected by its parent.

During testing, our executor attempts to concurrently access the parameters of the USB port while removing the USB hub (USB port's parent device) from the system. This vulnerability demonstrates that the relationship between devices affects the logic of data access between devices in the driver, as well as the parameter modification timing. Consequently, our mutation strategy successfully triggers a previously challenging-to-find vulnerability. Additionally, while this example illustrates how accessing device attributes can trigger the null pointer dereference through \texttt{usb\_hub\_to\_struct\_hub()}, it is important to note that this function is widely used in USB drivers and is not confined to the \texttt{sysfs} interface.

\section{Discussion}
In the following section, we will first discuss the severity of the bugs we discovered, and then discuss the limitations of our approach and point out the future research directions. 

\bheading{Bug severity.} 
Although the user needs root access to modify most of the runtime parameters, the severity of the discovered bugs is not trivial, for the following two reasons: 
\begin{packeditemize}
    \item First, some bugs are hidden by the default values which in other cases would not be the same. For example, in bug\#14 of \autoref{table5}, the bug cannot be triggered when shrinker of zswap is disabled, which is the case if CONFIG\_ZSWAP\_SHRINKER\_DEFAULT\_ON is not set during compilation. However, in other kernels that enable the config (i.e., Ubuntu enables it in the 24.04 LTS release), this bug can be triggered by unprivileged users.
    \item Second, we do find bugs, such as CVE-2024-36896, that can be triggered by reading parameters of the USB port, which is accessible by unprivileged users. 
\end{packeditemize}

Moreover, some parameters are frequently used by root-level tasks such as system daemon. Our work unveiled that if these parameters are not set properly by system managers, serious bugs could be triggered, such as deadlocks and use-after-free.

\bheading{Limitations and future work.} 
During the implementation of this work, we identified two limitations in our static analysis technique, including imprecision in analyzing indirect function calls and challenges in extracting valid attribute values. To address these issues, we plan to enhance the accuracy of identifying indirect call target functions by integrating dynamic information collection methods, such as ftrace, in future work. Additionally, we aim to improve the precision of parameter analysis by concolic execution.

\section{Related Work}

\bheading{Generic Kernel Fuzzing.} 
In recent years, considerable research efforts have focused on kernel fuzzing to enhance its performance across various domains. Google's Syzkaller stands out as the most widely used kernel fuzzer, enabling developers to continuously integrate syzlang to extend its fuzzing capabilities for testing new functional modules. For the problem that developing syzlang grammars is a manual process, and the number of supported devices and drivers is limited, recent advancements including DIFUZE\cite{corina2017difuze}, KSG\cite{sun2022ksg}, and SyzDescribe\cite{hao2023syzdescribe} use static analysis to recover the syscall interface syntax. Others leverage dynamic symbolic execution or even large language models to automatically generate syzlang from kernel source code\cite{chen2024syzgen++}\cite{yang2023kernelgpt}. 

Beyond syzlang generation, recent research has focused extensively on interrelationships between syzlangs~\cite{pailoor2018moonshine}\cite{wang2021syzvegas}. HEALER\cite{sun2021healer} investigates whether executing two syscalls together yields new code coverage compared to executing them separately, while ACTOR\cite{fleischer2023actor} scrutinizes the memory reads and writes associated with each syscall and synthesis use-after-free primitives.

Given that operating systems commonly run in multi-core environments, recent research has made progress in identifying concurrency vulnerabilities within the kernel~\cite{jeong2019razzer}\cite{yuan2023ddrace}\cite{jeong2024ozz}. For example, Krace\cite{xu2020krace} introduces a novel data contention metric, \emph{alias cov}, using it to direct fuzzing towards discovering new data contention groups. Building upon this, CONZZER\cite{jiang2022context} introduces a context-sensitive data contention fuzzing technique based on the observation of data race-triggering scenarios. Additionally, Segfuzz\cite{jeong2023segfuzz} explores the data contention results under different execution orders by mutating the execution order of kernel code fragments.

\bheading{Kernel Driver Fuzzing.} 
Due to the extensive volume of kernel driver code and the prevalence of vulnerabilities, significant research efforts have been dedicated to kernel driver fuzzing\cite{song2020agamotto}\cite{huster2024boldly}. PeriScope\cite{song2019periscope} and USBFuzz\cite{peng2020usbfuzz} inject interrupts at the network device and USB device level respectively to trigger the driver's execution. VIA\cite{hetzelt2021via} and PrIntFuzz\cite{ma2022printfuzz} create harnesses and perform fuzzing on the driver from the perspective of MMIO, PIO, and DMA. To address the issue of a limited number of simulated devices in QEMU, Drfuzz\cite{zhao2022semantic} and DevFuzz\cite{wu2023devfuzz} analyze device initialization processes, simulate interactions between devices and drivers, and enable fuzzing on device drivers in the absence of actual devices. In addition to focusing on open-source operating systems, there are several efforts to conduct fuzzing on drivers of closed-source operating systems\cite{chen2021syzgen}\cite{chen2024syzgen++}\cite{yin2023kextfuzz}.


\bheading{Static Analysis on Kernel Drivers.} 
In addition to fuzzing, static analysis methods are extensively employed in academia to identify vulnerabilities in operating systems. CRIX\cite{lu2019detecting} utilizes context-sensitive inter-procedural analysis to detect missing-check vulnerabilities in the kernel. To enhance the accuracy of static analysis, TyPM\cite{lu2023practical} filters out irrelevant modules based on parameter passing and inter-module dependencies, thereby reducing the scope of indirect call analysis and improving its accuracy. 
DCUAF\cite{bai2019effective} discovers concurrent UAF between driver threads, while UACatcher\cite{ma2023top} identifies a unique form of UAF vulnerability termed ``use after cleanup'', arising from a different order of resource release and resource use during device removal. PLA~\cite{ryan2023precise} integrates fuzzing and static analysis by initially capturing variable accesses and lock holdings. Subsequently, it calculates the likelihood of data contention between memory accesses based on sampled information to identify data contention issues within the kernel. LockPick\cite{cai2023place} and Archerfish\cite{yethreads} use static analysis and symbolic execution to solve locking problems in Linux.

\section{Conclusion}
In this paper, we introduce \sys, a novel kernel driver fuzzing framework that incorporates runtime parameters, including device attributes and kernel module parameters, into the fuzzing process. The core design of \sys is grounded in a comprehensive analysis of the driver model and its associated data structures. First, we propose an algorithm to extract parameters from the kernel source code along with their corresponding valid values to satisfy validation checks. Second, we establish multiple relationships to optimize the utilization of runtime parameters, including their associations with device files and the parameters of topologically connected devices. Third, we present a new mutation mechanism that integrates newly generated syzlangs into the existing fuzzing framework alongside pre-existing syzlangs. This mutation strategy encompasses both syzlang relationships and execution timing considerations. We evaluated \sys on the latest upstream kernel, uncovering 30 new bugs, with 14 being fixed upstream, and 9 of these receiving CVE assignments.

\bibliographystyle{ACM-Reference-Format}
\bibliography{ref}


\begin{thebibliography}{44}


\ifx \showCODEN    \undefined \def \showCODEN     #1{\unskip}     \fi
\ifx \showDOI      \undefined \def \showDOI       #1{#1}\fi
\ifx \showISBNx    \undefined \def \showISBNx     #1{\unskip}     \fi
\ifx \showISBNxiii \undefined \def \showISBNxiii  #1{\unskip}     \fi
\ifx \showISSN     \undefined \def \showISSN      #1{\unskip}     \fi
\ifx \showLCCN     \undefined \def \showLCCN      #1{\unskip}     \fi
\ifx \shownote     \undefined \def \shownote      #1{#1}          \fi
\ifx \showarticletitle \undefined \def \showarticletitle #1{#1}   \fi
\ifx \showURL      \undefined \def \showURL       {\relax}        \fi
\providecommand\bibfield[2]{#2}
\providecommand\bibinfo[2]{#2}
\providecommand\natexlab[1]{#1}
\providecommand\showeprint[2][]{arXiv:#2}

\bibitem[Bai et~al\mbox{.}(2019)]%
        {bai2019effective}
\bibfield{author}{\bibinfo{person}{Jia-Ju Bai}, \bibinfo{person}{Julia Lawall}, \bibinfo{person}{Qiu-Liang Chen}, {and} \bibinfo{person}{Shi-Min Hu}.} \bibinfo{year}{2019}\natexlab{}.
\newblock \showarticletitle{Effective static analysis of concurrency $\{$Use-After-Free$\}$ bugs in linux device drivers}. In \bibinfo{booktitle}{\emph{2019 USENIX Annual Technical Conference (USENIX ATC 19)}}. \bibinfo{pages}{255--268}.
\newblock


\bibitem[Bulekov et~al\mbox{.}(2023)]%
        {bulekov2023no}
\bibfield{author}{\bibinfo{person}{Alexander Bulekov}, \bibinfo{person}{Bandan Das}, \bibinfo{person}{Stefan Hajnoczi}, {and} \bibinfo{person}{Manuel Egele}.} \bibinfo{year}{2023}\natexlab{}.
\newblock \showarticletitle{No grammar, no problem: Towards fuzzing the linux kernel without system-call descriptions}. In \bibinfo{booktitle}{\emph{Network and Distributed System Security (NDSS) Symposium}}.
\newblock


\bibitem[Cai et~al\mbox{.}(2023)]%
        {cai2023place}
\bibfield{author}{\bibinfo{person}{Yuandao Cai}, \bibinfo{person}{Peisen Yao}, \bibinfo{person}{Chengfeng Ye}, {and} \bibinfo{person}{Charles Zhang}.} \bibinfo{year}{2023}\natexlab{}.
\newblock \showarticletitle{Place your locks well: understanding and detecting lock misuse bugs}. In \bibinfo{booktitle}{\emph{32nd USENIX Security Symposium (USENIX Security 23)}}. \bibinfo{pages}{3727--3744}.
\newblock


\bibitem[Chen et~al\mbox{.}(2024)]%
        {chen2024syzgen++}
\bibfield{author}{\bibinfo{person}{Weiteng Chen}, \bibinfo{person}{Yu Hao}, \bibinfo{person}{Zheng Zhang}, \bibinfo{person}{Xiaochen Zou}, \bibinfo{person}{Dhilung Kirat}, \bibinfo{person}{Shachee Mishra}, \bibinfo{person}{Douglas Schales}, \bibinfo{person}{Jiyong Jang}, {and} \bibinfo{person}{Zhiyun Qian}.} \bibinfo{year}{2024}\natexlab{}.
\newblock \showarticletitle{SyzGen++: Dependency Inference for Augmenting Kernel Driver Fuzzing}. In \bibinfo{booktitle}{\emph{IEEE Symposium on Security and Privacy}}.
\newblock


\bibitem[Chen et~al\mbox{.}(2021)]%
        {chen2021syzgen}
\bibfield{author}{\bibinfo{person}{Weiteng Chen}, \bibinfo{person}{Yu Wang}, \bibinfo{person}{Zheng Zhang}, {and} \bibinfo{person}{Zhiyun Qian}.} \bibinfo{year}{2021}\natexlab{}.
\newblock \showarticletitle{Syzgen: Automated generation of syscall specification of closed-source macos drivers}. In \bibinfo{booktitle}{\emph{Proceedings of the 2021 ACM SIGSAC Conference on Computer and Communications Security}}. \bibinfo{pages}{749--763}.
\newblock


\bibitem[Corina et~al\mbox{.}(2017)]%
        {corina2017difuze}
\bibfield{author}{\bibinfo{person}{Jake Corina}, \bibinfo{person}{Aravind Machiry}, \bibinfo{person}{Christopher Salls}, \bibinfo{person}{Yan Shoshitaishvili}, \bibinfo{person}{Shuang Hao}, \bibinfo{person}{Christopher Kruegel}, {and} \bibinfo{person}{Giovanni Vigna}.} \bibinfo{year}{2017}\natexlab{}.
\newblock \showarticletitle{Difuze: Interface aware fuzzing for kernel drivers}. In \bibinfo{booktitle}{\emph{Proceedings of the 2017 ACM SIGSAC Conference on Computer and Communications Security}}. \bibinfo{pages}{2123--2138}.
\newblock


\bibitem[Fleischer et~al\mbox{.}(2023)]%
        {fleischer2023actor}
\bibfield{author}{\bibinfo{person}{Marius Fleischer}, \bibinfo{person}{Dipanjan Das}, \bibinfo{person}{Priyanka Bose}, \bibinfo{person}{Weiheng Bai}, \bibinfo{person}{Kangjie Lu}, \bibinfo{person}{Mathias Payer}, \bibinfo{person}{Christopher Kruegel}, {and} \bibinfo{person}{Giovanni Vigna}.} \bibinfo{year}{2023}\natexlab{}.
\newblock \showarticletitle{$\{$ACTOR$\}$:$\{$Action-Guided$\}$ Kernel Fuzzing}. In \bibinfo{booktitle}{\emph{32nd USENIX Security Symposium (USENIX Security 23)}}. \bibinfo{pages}{5003--5020}.
\newblock


\bibitem[Google(2000)]%
        {LLVM}
\bibfield{author}{\bibinfo{person}{Google}.} \bibinfo{year}{2000}\natexlab{}.
\newblock \bibinfo{title}{The LLVM Compiler Infrastructure}.
\newblock \bibinfo{howpublished}{\url{https://llvm.org/}}.
\newblock


\bibitem[Google(2015a)]%
        {syzlang}
\bibfield{author}{\bibinfo{person}{Google}.} \bibinfo{year}{2015}\natexlab{a}.
\newblock \bibinfo{title}{Syscall description language}.
\newblock \bibinfo{howpublished}{\url{https://github.com/google/syzkaller/blob/master/docs/syscall_descriptions_syntax.md}}.
\newblock


\bibitem[Google(2015b)]%
        {Syzkaller}
\bibfield{author}{\bibinfo{person}{Google}.} \bibinfo{year}{2015}\natexlab{b}.
\newblock \bibinfo{title}{Syzkaller}.
\newblock \bibinfo{howpublished}{\url{https://github.com/google/syzkaller}}.
\newblock


\bibitem[Hao et~al\mbox{.}(2023)]%
        {hao2023syzdescribe}
\bibfield{author}{\bibinfo{person}{Yu Hao}, \bibinfo{person}{Guoren Li}, \bibinfo{person}{Xiaochen Zou}, \bibinfo{person}{Weiteng Chen}, \bibinfo{person}{Shitong Zhu}, \bibinfo{person}{Zhiyun Qian}, {and} \bibinfo{person}{Ardalan~Amiri Sani}.} \bibinfo{year}{2023}\natexlab{}.
\newblock \showarticletitle{Syzdescribe: Principled, automated, static generation of syscall descriptions for kernel drivers}. In \bibinfo{booktitle}{\emph{2023 IEEE Symposium on Security and Privacy (SP)}}. IEEE, \bibinfo{pages}{3262--3278}.
\newblock


\bibitem[Hetzelt et~al\mbox{.}(2021)]%
        {hetzelt2021via}
\bibfield{author}{\bibinfo{person}{Felicitas Hetzelt}, \bibinfo{person}{Martin Radev}, \bibinfo{person}{Robert Buhren}, \bibinfo{person}{Mathias Morbitzer}, {and} \bibinfo{person}{Jean-Pierre Seifert}.} \bibinfo{year}{2021}\natexlab{}.
\newblock \showarticletitle{Via: Analyzing device interfaces of protected virtual machines}. In \bibinfo{booktitle}{\emph{Proceedings of the 37th Annual Computer Security Applications Conference}}. \bibinfo{pages}{273--284}.
\newblock


\bibitem[Huster et~al\mbox{.}(2024)]%
        {huster2024boldly}
\bibfield{author}{\bibinfo{person}{S{\"o}nke Huster}, \bibinfo{person}{Matthias Hollick}, {and} \bibinfo{person}{Jiska Classen}.} \bibinfo{year}{2024}\natexlab{}.
\newblock \showarticletitle{To boldly go where no fuzzer has gone before: Finding bugs in linux’wireless stacks through virtio devices}. In \bibinfo{booktitle}{\emph{2024 IEEE Symposium on Security and Privacy (SP). Los Alamitos, CA, USA: IEEE Computer Society}}. \bibinfo{pages}{24--24}.
\newblock


\bibitem[Jeong et~al\mbox{.}(2024)]%
        {jeong2024ozz}
\bibfield{author}{\bibinfo{person}{Dae~R Jeong}, \bibinfo{person}{Yewon Choi}, \bibinfo{person}{Byoungyoung Lee}, \bibinfo{person}{Insik Shin}, {and} \bibinfo{person}{Youngjin Kwon}.} \bibinfo{year}{2024}\natexlab{}.
\newblock \showarticletitle{OZZ: Identifying Kernel Out-of-Order Concurrency Bugs with In-Vivo Memory Access Reordering}. In \bibinfo{booktitle}{\emph{Proceedings of the ACM SIGOPS 30th Symposium on Operating Systems Principles}}. \bibinfo{pages}{229--248}.
\newblock


\bibitem[Jeong et~al\mbox{.}(2019)]%
        {jeong2019razzer}
\bibfield{author}{\bibinfo{person}{Dae~R Jeong}, \bibinfo{person}{Kyungtae Kim}, \bibinfo{person}{Basavesh Shivakumar}, \bibinfo{person}{Byoungyoung Lee}, {and} \bibinfo{person}{Insik Shin}.} \bibinfo{year}{2019}\natexlab{}.
\newblock \showarticletitle{Razzer: Finding kernel race bugs through fuzzing}. In \bibinfo{booktitle}{\emph{2019 IEEE Symposium on Security and Privacy (SP)}}. IEEE, \bibinfo{pages}{754--768}.
\newblock


\bibitem[Jeong et~al\mbox{.}(2023)]%
        {jeong2023segfuzz}
\bibfield{author}{\bibinfo{person}{Dae~R Jeong}, \bibinfo{person}{Byoungyoung Lee}, \bibinfo{person}{Insik Shin}, {and} \bibinfo{person}{Youngjin Kwon}.} \bibinfo{year}{2023}\natexlab{}.
\newblock \showarticletitle{Segfuzz: Segmentizing thread interleaving to discover kernel concurrency bugs through fuzzing}. In \bibinfo{booktitle}{\emph{2023 IEEE Symposium on Security and Privacy (SP)}}. IEEE, \bibinfo{pages}{2104--2121}.
\newblock


\bibitem[Jiang et~al\mbox{.}(2022)]%
        {jiang2022context}
\bibfield{author}{\bibinfo{person}{Zu-Ming Jiang}, \bibinfo{person}{Jia-Ju Bai}, \bibinfo{person}{Kangjie Lu}, {and} \bibinfo{person}{Shi-Min Hu}.} \bibinfo{year}{2022}\natexlab{}.
\newblock \showarticletitle{Context-sensitive and directional concurrency fuzzing for data-race detection}. In \bibinfo{booktitle}{\emph{Network and Distributed Systems Security (NDSS) Symposium 2022}}.
\newblock


\bibitem[Kadav and Swift(2012)]%
        {kadav2012understanding}
\bibfield{author}{\bibinfo{person}{Asim Kadav} {and} \bibinfo{person}{Michael~M Swift}.} \bibinfo{year}{2012}\natexlab{}.
\newblock \showarticletitle{Understanding modern device drivers}.
\newblock \bibinfo{journal}{\emph{ACM SIGPLAN Notices}} \bibinfo{volume}{47}, \bibinfo{number}{4} (\bibinfo{year}{2012}), \bibinfo{pages}{87--98}.
\newblock


\bibitem[Kim et~al\mbox{.}(2022)]%
        {kim2022fuzzusb}
\bibfield{author}{\bibinfo{person}{Kyungtae Kim}, \bibinfo{person}{Taegyu Kim}, \bibinfo{person}{Ertza Warraich}, \bibinfo{person}{Byoungyoung Lee}, \bibinfo{person}{Kevin~RB Butler}, \bibinfo{person}{Antonio Bianchi}, {and} \bibinfo{person}{Dave~Jing Tian}.} \bibinfo{year}{2022}\natexlab{}.
\newblock \showarticletitle{Fuzzusb: Hybrid stateful fuzzing of usb gadget stacks}. In \bibinfo{booktitle}{\emph{2022 IEEE Symposium on Security and Privacy (SP)}}. IEEE, \bibinfo{pages}{2212--2229}.
\newblock


\bibitem[Lu(2023)]%
        {lu2023practical}
\bibfield{author}{\bibinfo{person}{Kangjie Lu}.} \bibinfo{year}{2023}\natexlab{}.
\newblock \showarticletitle{Practical program modularization with type-based dependence analysis}. In \bibinfo{booktitle}{\emph{2023 IEEE Symposium on Security and Privacy (SP)}}. IEEE, \bibinfo{pages}{1256--1270}.
\newblock


\bibitem[Lu et~al\mbox{.}(2019)]%
        {lu2019detecting}
\bibfield{author}{\bibinfo{person}{Kangjie Lu}, \bibinfo{person}{Aditya Pakki}, {and} \bibinfo{person}{Qiushi Wu}.} \bibinfo{year}{2019}\natexlab{}.
\newblock \showarticletitle{Detecting $\{$Missing-Check$\}$ bugs via semantic-and $\{$Context-Aware$\}$ criticalness and constraints inferences}. In \bibinfo{booktitle}{\emph{28th USENIX Security Symposium (USENIX Security 19)}}. \bibinfo{pages}{1769--1786}.
\newblock


\bibitem[Ma et~al\mbox{.}(2023)]%
        {ma2023top}
\bibfield{author}{\bibinfo{person}{Lin Ma}, \bibinfo{person}{Duoming Zhou}, \bibinfo{person}{Hanjie Wu}, \bibinfo{person}{Yajin Zhou}, \bibinfo{person}{Rui Chang}, \bibinfo{person}{Hao Xiong}, \bibinfo{person}{Lei Wu}, {and} \bibinfo{person}{Kui Ren}.} \bibinfo{year}{2023}\natexlab{}.
\newblock \showarticletitle{When Top-down Meets Bottom-up: Detecting and Exploiting Use-After-Cleanup Bugs in Linux Kernel}. In \bibinfo{booktitle}{\emph{2023 IEEE Symposium on Security and Privacy (SP)}}. IEEE, \bibinfo{pages}{2138--2154}.
\newblock


\bibitem[Ma et~al\mbox{.}(2022)]%
        {ma2022printfuzz}
\bibfield{author}{\bibinfo{person}{Zheyu Ma}, \bibinfo{person}{Bodong Zhao}, \bibinfo{person}{Letu Ren}, \bibinfo{person}{Zheming Li}, \bibinfo{person}{Siqi Ma}, \bibinfo{person}{Xiapu Luo}, {and} \bibinfo{person}{Chao Zhang}.} \bibinfo{year}{2022}\natexlab{}.
\newblock \showarticletitle{Printfuzz: Fuzzing linux drivers via automated virtual device simulation}. In \bibinfo{booktitle}{\emph{Proceedings of the 31st ACM SIGSOFT International Symposium on Software Testing and Analysis}}. \bibinfo{pages}{404--416}.
\newblock


\bibitem[Mochel(2002)]%
        {mochel2002linux}
\bibfield{author}{\bibinfo{person}{Patrick Mochel}.} \bibinfo{year}{2002}\natexlab{}.
\newblock \showarticletitle{The linux kernel device model}. In \bibinfo{booktitle}{\emph{Ottawa Linux Symposium}}. \bibinfo{pages}{368}.
\newblock


\bibitem[Pailoor et~al\mbox{.}(2018)]%
        {pailoor2018moonshine}
\bibfield{author}{\bibinfo{person}{Shankara Pailoor}, \bibinfo{person}{Andrew Aday}, {and} \bibinfo{person}{Suman Jana}.} \bibinfo{year}{2018}\natexlab{}.
\newblock \showarticletitle{$\{$MoonShine$\}$: Optimizing $\{$OS$\}$ fuzzer seed selection with trace distillation}. In \bibinfo{booktitle}{\emph{27th USENIX Security Symposium (USENIX Security 18)}}. \bibinfo{pages}{729--743}.
\newblock


\bibitem[Peng and Payer(2020)]%
        {peng2020usbfuzz}
\bibfield{author}{\bibinfo{person}{Hui Peng} {and} \bibinfo{person}{Mathias Payer}.} \bibinfo{year}{2020}\natexlab{}.
\newblock \showarticletitle{$\{$USBFuzz$\}$: A Framework for Fuzzing $\{$USB$\}$ Drivers by Device Emulation}. In \bibinfo{booktitle}{\emph{29th USENIX Security Symposium (USENIX Security 20)}}. \bibinfo{pages}{2559--2575}.
\newblock


\bibitem[Ryan et~al\mbox{.}(2023)]%
        {ryan2023precise}
\bibfield{author}{\bibinfo{person}{Gabriel Ryan}, \bibinfo{person}{Abhishek Shah}, \bibinfo{person}{Dongdong She}, {and} \bibinfo{person}{Suman Jana}.} \bibinfo{year}{2023}\natexlab{}.
\newblock \showarticletitle{Precise Detection of Kernel Data Races with Probabilistic Lockset Analysis}. In \bibinfo{booktitle}{\emph{2023 IEEE Symposium on Security and Privacy (SP)}}. IEEE, \bibinfo{pages}{2086--2103}.
\newblock


\bibitem[Schumilo et~al\mbox{.}(2017)]%
        {schumilo2017kafl}
\bibfield{author}{\bibinfo{person}{Sergej Schumilo}, \bibinfo{person}{Cornelius Aschermann}, \bibinfo{person}{Robert Gawlik}, \bibinfo{person}{Sebastian Schinzel}, {and} \bibinfo{person}{Thorsten Holz}.} \bibinfo{year}{2017}\natexlab{}.
\newblock \showarticletitle{$\{$kAFL$\}$:$\{$Hardware-Assisted$\}$ feedback fuzzing for $\{$OS$\}$ kernels}. In \bibinfo{booktitle}{\emph{26th USENIX security symposium (USENIX Security 17)}}. \bibinfo{pages}{167--182}.
\newblock


\bibitem[Shameli-Sendi(2021)]%
        {shameli2021understanding}
\bibfield{author}{\bibinfo{person}{Alireza Shameli-Sendi}.} \bibinfo{year}{2021}\natexlab{}.
\newblock \showarticletitle{Understanding Linux kernel vulnerabilities}.
\newblock \bibinfo{journal}{\emph{Journal of Computer Virology and Hacking Techniques}} \bibinfo{volume}{17}, \bibinfo{number}{4} (\bibinfo{year}{2021}), \bibinfo{pages}{265--278}.
\newblock


\bibitem[Shen et~al\mbox{.}(2022)]%
        {shen2022drifuzz}
\bibfield{author}{\bibinfo{person}{Zekun Shen}, \bibinfo{person}{Ritik Roongta}, {and} \bibinfo{person}{Brendan Dolan-Gavitt}.} \bibinfo{year}{2022}\natexlab{}.
\newblock \showarticletitle{Drifuzz: Harvesting bugs in device drivers from golden seeds}. In \bibinfo{booktitle}{\emph{31st USENIX Security Symposium (USENIX Security 22)}}. \bibinfo{pages}{1275--1290}.
\newblock


\bibitem[Song et~al\mbox{.}(2019)]%
        {song2019periscope}
\bibfield{author}{\bibinfo{person}{Dokyung Song}, \bibinfo{person}{Felicitas Hetzelt}, \bibinfo{person}{Dipanjan Das}, \bibinfo{person}{Chad Spensky}, \bibinfo{person}{Yeoul Na}, \bibinfo{person}{Stijn Volckaert}, \bibinfo{person}{Giovanni Vigna}, \bibinfo{person}{Christopher Kruegel}, \bibinfo{person}{Jean-Pierre Seifert}, {and} \bibinfo{person}{Michael Franz}.} \bibinfo{year}{2019}\natexlab{}.
\newblock \showarticletitle{Periscope: An effective probing and fuzzing framework for the hardware-os boundary}. In \bibinfo{booktitle}{\emph{NDSS}}.
\newblock


\bibitem[Song et~al\mbox{.}(2020)]%
        {song2020agamotto}
\bibfield{author}{\bibinfo{person}{Dokyung Song}, \bibinfo{person}{Felicitas Hetzelt}, \bibinfo{person}{Jonghwan Kim}, \bibinfo{person}{Brent~Byunghoon Kang}, \bibinfo{person}{Jean-Pierre Seifert}, {and} \bibinfo{person}{Michael Franz}.} \bibinfo{year}{2020}\natexlab{}.
\newblock \showarticletitle{Agamotto: Accelerating kernel driver fuzzing with lightweight virtual machine checkpoints}. In \bibinfo{booktitle}{\emph{29th USENIX Security Symposium (USENIX Security 20)}}. \bibinfo{pages}{2541--2557}.
\newblock


\bibitem[Sun et~al\mbox{.}(2022)]%
        {sun2022ksg}
\bibfield{author}{\bibinfo{person}{Hao Sun}, \bibinfo{person}{Yuheng Shen}, \bibinfo{person}{Jianzhong Liu}, \bibinfo{person}{Yiru Xu}, {and} \bibinfo{person}{Yu Jiang}.} \bibinfo{year}{2022}\natexlab{}.
\newblock \showarticletitle{$\{$KSG$\}$: Augmenting kernel fuzzing with system call specification generation}. In \bibinfo{booktitle}{\emph{2022 USENIX Annual Technical Conference (USENIX ATC 22)}}. \bibinfo{pages}{351--366}.
\newblock


\bibitem[Sun et~al\mbox{.}(2021)]%
        {sun2021healer}
\bibfield{author}{\bibinfo{person}{Hao Sun}, \bibinfo{person}{Yuheng Shen}, \bibinfo{person}{Cong Wang}, \bibinfo{person}{Jianzhong Liu}, \bibinfo{person}{Yu Jiang}, \bibinfo{person}{Ting Chen}, {and} \bibinfo{person}{Aiguo Cui}.} \bibinfo{year}{2021}\natexlab{}.
\newblock \showarticletitle{Healer: Relation learning guided kernel fuzzing}. In \bibinfo{booktitle}{\emph{Proceedings of the ACM SIGOPS 28th Symposium on Operating Systems Principles}}. \bibinfo{pages}{344--358}.
\newblock


\bibitem[Wang et~al\mbox{.}(2021)]%
        {wang2021syzvegas}
\bibfield{author}{\bibinfo{person}{Daimeng Wang}, \bibinfo{person}{Zheng Zhang}, \bibinfo{person}{Hang Zhang}, \bibinfo{person}{Zhiyun Qian}, \bibinfo{person}{Srikanth~V Krishnamurthy}, {and} \bibinfo{person}{Nael Abu-Ghazaleh}.} \bibinfo{year}{2021}\natexlab{}.
\newblock \showarticletitle{$\{$SyzVegas$\}$: Beating kernel fuzzing odds with reinforcement learning}. In \bibinfo{booktitle}{\emph{30th USENIX Security Symposium (USENIX Security 21)}}. \bibinfo{pages}{2741--2758}.
\newblock


\bibitem[Wu et~al\mbox{.}(2023)]%
        {wu2023devfuzz}
\bibfield{author}{\bibinfo{person}{Yilun Wu}, \bibinfo{person}{Tong Zhang}, \bibinfo{person}{Changhee Jung}, {and} \bibinfo{person}{Dongyoon Lee}.} \bibinfo{year}{2023}\natexlab{}.
\newblock \showarticletitle{DEVFUZZ: automatic device model-guided device driver fuzzing}. In \bibinfo{booktitle}{\emph{2023 IEEE Symposium on Security and Privacy (SP)}}. IEEE, \bibinfo{pages}{3246--3261}.
\newblock


\bibitem[Xu et~al\mbox{.}(2024)]%
        {xumock}
\bibfield{author}{\bibinfo{person}{Jiacheng Xu}, \bibinfo{person}{Xuhong Zhang}, \bibinfo{person}{Shouling Ji}, \bibinfo{person}{Yuan Tian}, \bibinfo{person}{Binbin Zhao}, \bibinfo{person}{Qinying Wang}, \bibinfo{person}{Peng Cheng}, {and} \bibinfo{person}{Jiming Chen}.} \bibinfo{year}{2024}\natexlab{}.
\newblock \showarticletitle{MOCK: Optimizing Kernel Fuzzing Mutation with Context-aware Dependency}.
\newblock


\bibitem[Xu et~al\mbox{.}(2020)]%
        {xu2020krace}
\bibfield{author}{\bibinfo{person}{Meng Xu}, \bibinfo{person}{Sanidhya Kashyap}, \bibinfo{person}{Hanqing Zhao}, {and} \bibinfo{person}{Taesoo Kim}.} \bibinfo{year}{2020}\natexlab{}.
\newblock \showarticletitle{Krace: Data race fuzzing for kernel file systems}. In \bibinfo{booktitle}{\emph{2020 IEEE Symposium on Security and Privacy (SP)}}. IEEE, \bibinfo{pages}{1643--1660}.
\newblock


\bibitem[Yang et~al\mbox{.}(2023)]%
        {yang2023kernelgpt}
\bibfield{author}{\bibinfo{person}{Chenyuan Yang}, \bibinfo{person}{Zijie Zhao}, {and} \bibinfo{person}{Lingming Zhang}.} \bibinfo{year}{2023}\natexlab{}.
\newblock \showarticletitle{Kernelgpt: Enhanced kernel fuzzing via large language models}.
\newblock \bibinfo{journal}{\emph{arXiv preprint arXiv:2401.00563}} (\bibinfo{year}{2023}).
\newblock


\bibitem[Ye et~al\mbox{.}(2024)]%
        {yethreads}
\bibfield{author}{\bibinfo{person}{Chengfeng Ye}, \bibinfo{person}{Yuandao Cai}, {and} \bibinfo{person}{Charles Zhang}.} \bibinfo{year}{2024}\natexlab{}.
\newblock \showarticletitle{When Threads Meet Interrupts: Effective Static Detection of Interrupt-Based Deadlocks in Linux}. In \bibinfo{booktitle}{\emph{33st USENIX Security Symposium (USENIX Security 24)}}.
\newblock


\bibitem[Yin et~al\mbox{.}(2023)]%
        {yin2023kextfuzz}
\bibfield{author}{\bibinfo{person}{Tingting Yin}, \bibinfo{person}{Zicong Gao}, \bibinfo{person}{Zhenghang Xiao}, \bibinfo{person}{Zheyu Ma}, \bibinfo{person}{Min Zheng}, {and} \bibinfo{person}{Chao Zhang}.} \bibinfo{year}{2023}\natexlab{}.
\newblock \showarticletitle{$\{$KextFuzz$\}$: Fuzzing $\{$macOS$\}$ Kernel $\{$EXTensions$\}$ on Apple Silicon via Exploiting Mitigations}. In \bibinfo{booktitle}{\emph{32nd USENIX Security Symposium (USENIX Security 23)}}. \bibinfo{pages}{5039--5054}.
\newblock


\bibitem[Yuan et~al\mbox{.}(2023)]%
        {yuan2023ddrace}
\bibfield{author}{\bibinfo{person}{Ming Yuan}, \bibinfo{person}{Bodong Zhao}, \bibinfo{person}{Penghui Li}, \bibinfo{person}{Jiashuo Liang}, \bibinfo{person}{Xinhui Han}, \bibinfo{person}{Xiapu Luo}, {and} \bibinfo{person}{Chao Zhang}.} \bibinfo{year}{2023}\natexlab{}.
\newblock \showarticletitle{DDRace: Finding Concurrency UAF Vulnerabilities in Linux Drivers with Directed Fuzzing.}. In \bibinfo{booktitle}{\emph{USENIX Security Symposium}}. \bibinfo{pages}{2849--2866}.
\newblock


\bibitem[Zhao et~al\mbox{.}(2022a)]%
        {zhao2022statefuzz}
\bibfield{author}{\bibinfo{person}{Bodong Zhao}, \bibinfo{person}{Zheming Li}, \bibinfo{person}{Shisong Qin}, \bibinfo{person}{Zheyu Ma}, \bibinfo{person}{Ming Yuan}, \bibinfo{person}{Wenyu Zhu}, \bibinfo{person}{Zhihong Tian}, {and} \bibinfo{person}{Chao Zhang}.} \bibinfo{year}{2022}\natexlab{a}.
\newblock \showarticletitle{$\{$StateFuzz$\}$: System $\{$Call-Based$\}$$\{$State-Aware$\}$ Linux Driver Fuzzing}. In \bibinfo{booktitle}{\emph{31st USENIX Security Symposium (USENIX Security 22)}}. \bibinfo{pages}{3273--3289}.
\newblock


\bibitem[Zhao et~al\mbox{.}(2022b)]%
        {zhao2022semantic}
\bibfield{author}{\bibinfo{person}{Wenjia Zhao}, \bibinfo{person}{Kangjie Lu}, \bibinfo{person}{Qiushi Wu}, {and} \bibinfo{person}{Yong Qi}.} \bibinfo{year}{2022}\natexlab{b}.
\newblock \showarticletitle{Semantic-informed driver fuzzing without both the hardware devices and the emulators}. In \bibinfo{booktitle}{\emph{Network and Distributed Systems Security (NDSS) Symposium 2022}}.
\newblock


\end{thebibliography}
\appendix
\begin{table}[t]
\centering
\caption{Number of enabled syzlang, device attributes, and kernel module parameters of each driver.}
\resizebox{\columnwidth}{!}{
\begin{tabular}{lccc} 
\toprule
\textbf{Driver}            & \textbf{Syzlang} & \textbf{Module Param.} & \textbf{Device Attr.}  \\
\hline
seq               &    32   &    9    &   11   \\
usb               &    30   &    0    &   15   \\
mouse             &    3    &    18   &   2    \\
sr                &    43   &    0    &   32   \\
msr               &    3    &    3    &   3    \\
loop              &    15   &    0    &   24   \\
i2c               &    10   &    4    &   6    \\
rtc               &    23   &    0    &   6    \\

\hline
\end{tabular}}
\label{param_num}
\end{table}

\section{Statistics of Influenced Control Flow}\label{exp_setup}

We did a static taint flow analysis to quantify the code dominated by runtime parameters and syscall arguments. Our experiment is done on Linux upstream v6.10, with all-yes-config. We use CodeQL as our analysis tool. We first demonstrate how we identify taint sources and then illustrate the taint analysis details.

\bheading{Kernel module parameters collection.}
Since the Linux kernel uses ``\emph{\_\_module\_param\_call}'' macro to export all kernel module parameters to files in userspace, we check all the invoking sites of this macro and identify corresponding kernel module parameters. We only collect writable parameters since read-only parameters cannot contribute to coverage during fuzzing.

\bheading{Device attributes collection.}
Device attributes are created and initialized during device registers. Kernel developers use ``\emph{struct device\_attribute}'' to define show/store interfaces, and target variables could be found through these functions. ``show'' functions export variable values as strings to userspace. In contrast, ``store'' functions take user string as input and modify kernel variables accordingly. Specifically, we first identify these interface structures and filter out all the structures without the ``store'' field for the same reason above. In ``show'' functions, device attribute values are transferred to string through ``\emph{sysfs\_emit()}'' or ``\emph{\%printf()}'' (including sprintf, snprintf, scnprintf, etc.). We count and find out that over 82.6\% ``show'' functions follow this design pattern. Thus, we can identify device attribute variables.

\bheading{Syscall arguments collection.}
To make the statistics more persuasive, we also compare the arguments of syscalls since they are also controllable by users. However, system calls will finally be dispatched to drivers through indirect function calls, which are hard to track. Inspired by \cite{hao2023syzdescribe}, callback functions are registered during driver initialization and can be found in ``struct file\_operations'' for character devices and ``struct block\_device\_operations'' for block devices. Specifically, network devices use ``struct net\_device\_ops'' to handle network callbacks. We filter commonly use functions including ``open'', ``read'', ``write'', ``ioctl'', ``mmap'', ``llseek'' and ``release''. All the arguments of those functions are regarded as controllable by users. We found 2,685 global variables containing callback function pointers, with 4,548 target functions and 12,839 arguments.

\bheading{Taint analysis.}
We conduct an inter-procedural field-sensitive taint analysis, using the above three as taint sources. Since some user arguments are pointers to user space memory, which need ``\emph{copy\_from\_ user()}'' and ``\emph{get\_user()}'' to get a copy in kernel space, we add additional flow steps to mark the \textbf{dst} argument as tainted accordingly. Also, the \textbf{dst} target may be used as a pointer. For example, when calling ``\emph{copy\_from\_user}(\textbf{kernel\_buf}, user\_buf, count)'', the ``\textbf{kernel\_buf}'' is a pointer pointing to kernel memory, which will later be used during subfield access such as ``\textbf{kernel\_buf}$\rightarrow$subfield'', and the subfield is also a taint source. We track the usage of taint sources and mark the target of pointer-field access as tainted. 

\section{Number of runtime parameters}
The number of syzlangs, device attributes, and kernel module parameters of each driver. Note that not all drivers have kernel module parameters. The list shows that even a limited number of runtime parameters can influence code execution significantly. For instance, driver \emph{msr} increases the coverage by 34\% with only 6 parameters.

\end{document}